\documentclass{article}
 \usepackage{amssymb,amsmath,graphicx}
 \numberwithin{equation}{section}
 \title{Law of Excluded Quantum Gambling Strategies}
 \author{Gavriel  Segre \thanks{I want strongly to thank Fabio Benatti form many invaluable remarks, suggestions and improvements}}
 \author{Universit\'{a} di Pavia, Italia}
 \newtheorem{definition}{Definition}[section]
 \newtheorem{theorem}{Theorem}[section]
 \newtheorem{trial definition}{Trial definition}[section]
 
 \newtheorem{axiom}{AXIOM}[section]
 \newenvironment{hypothesis}{\newline HP:} {\newline}
 \newenvironment{thesis}{\newline TH:} {\newline}
 \newenvironment{proof}{\begin{center}PROOF: \end{center}} {
$ \blacksquare $}
 \newtheorem{conjecture}{Conjecture}[section]
 \newtheorem{example}{Example}[section]
 \begin{document}
 \maketitle
 \begin{abstract}
We introduce and analyze a quantum analogue of the Law of Excluded
Gambling Strategies of Classical Decision Theory by the definition
of different kind of quantum casinos.

The necessity of keeping into account entaglement (by the way we
give a staightforward generalization of Schmidt's entanglement
measure) forces us to adopt the general algebraic language of
Quantum Probability Theory whose essential points are reviewed.

The Mathematica code of two packages simulating, respectively,
classical and quantum gambling is included.

The deep link  existing between the censorship of winning quantum
gambling strategies and the central notion of Quantum Algorithmic
Information Theory, namely quantum algorithmic randomness (by the
way we introduce and discard the naive noncommutative
generalization of the original Kolmogorov definition), is analyzed
 \end{abstract}
 \newpage
 \tableofcontents
 \newpage
 \section{Von Mises' Frequentistic Foundation of Probability}
The mirable features of the Kolmogorovian measure-theoretic
axiomatization of  Classical Probability Theory
\cite{Kolmogorov-56} has lead to consider it as the last word
about Foundations of Classical Probability Theory, leading to the
general attitude of forgetting the other different
axiomatizations and, in particular, von Mises' Frequentistic one
\cite{von-Mises-81}.

Richard Von Mises' axiomatization of Classical Probability Theory
lies on the mathematical formalization of the following two
empirical laws:
\begin{enumerate}
  \item \textbf{Law of Stability of Statistic Relative
  Frequencies}
\begin{quote}
It is essential for the theory of probability that experience has
shown that in the game of dice, as in all other mass phenomena
which we have mentioned, the relative frequencies of certain
attributes become more and more stable as the number of
observations is increased (cfr. pag.12 of \cite{von-Mises-81})
\end{quote}

  \item \textbf{Law of Excluded Gambling Strategies}
\begin{quote}
Everybody who has been to Monte Carlo, or who has read
descriptions of a gambling bank, know how many 'absolutely safe'
gambling systems, sometimes of an enormously complicated
character, have been invented and tried out by gamblers; and new
systems are still suggested every day. The authors of such
systems have all, sooner or later, had the sad experience of
finding out that no system is able to improve their chance of
winning in the long run,i.e. to affect the relative frequencies
with which different colours of numbers appear in a sequence
selected from the total sequence of the game. This experience
forms the experimental basis of our definition of probability.
(cfr. pagg.25-26 of \cite{von-Mises-81})
\end{quote}
\end{enumerate}
According to Von Mises Probability Theory concerns properties of
collectivities, i.e. of sequences of identical objects.

Considering each individual object as a letter of an alphabet $
\Sigma$, we can then say that Probability Theory concerns
elements of the set $ \Sigma^{\infty} $ of the sequences of
letters from $ \Sigma $ or, more properly, a certain subset  $
{\mathcal{C}}ollectives\; \subset \;  \Sigma^{\infty} $ whose
elements are called \textbf{collectives}, where:
\begin{definition} \label{def: classical strings on an alphabet}
\end{definition}
SET OF THE STRINGS ON $ \Sigma $ :
\begin{equation}
\Sigma^{\star} \; \equiv \; \cup_{ k \in {\mathbb{N}}} \Sigma^{k}
\end{equation}
\begin{definition} \label{def: classical sequences on an alphabet}
\end{definition}
SET OF THE SEQUENCES ON $ \Sigma $ :
\begin{equation}
\Sigma^{\infty} \; \equiv \; \{ \lambda \} \; \bigcup \;\{ \bar{x}
: {\mathbb{N}}_{+} \, \rightarrow \, \Sigma \}
\end{equation}
where $ \lambda $ denotes the \textit{empty string}.

Given $ \vec{x} \in \Sigma^{\star} $ let us denote by $
\vec{x}^{n} \in \Sigma^{\star} $ the string made of n repetitions
of $ \vec{x} $ and by $ a^{\infty} \in \Sigma^{\infty} $ the
sequence made of infinite repetitions of $ \vec{x} $.

It is important to remark that \cite{Calude-94}:
\begin{theorem} \label{th:cardinalities of strings and sequences}
\end{theorem}
ON THE CARDINALITIES OF STRINGS AND SEQUENCES OVER A FINITE
ALPHABET
\begin{hypothesis}
\end{hypothesis}
\begin{equation*}
  cardinality ( \Sigma ) \; \in \; {\mathbb{N}}
\end{equation*}
\begin{thesis}
\end{thesis}
\begin{align*}
  cardinality(\Sigma^{\star}) \; & = \; \aleph_{0}   \\
  cardinality(\Sigma^{\infty}) \; & = \;  \aleph_{1}
\end{align*}
Let us then introduce the set $ {{\mathcal{A}}ttributes} ( \Sigma
) $ of the \textbf{attributes} of $ \mathcal{C} $'s elements
defined as the set of unary predicates about the generic $ C \;
\in  \; {\mathcal{C}}ollectives $.

The mathematical formalization of the \textbf{Law of Stability of
Statistic Relative Frequencies} results in the following:
\begin{axiom} \label{ax:axiom of convergence}
\end{axiom}
AXIOM OF CONVERGENCE
\begin{hypothesis}
\end{hypothesis}
\begin{equation*}
   C \; \in \; {\mathcal{C}}ollectives
\end{equation*}
\begin{equation*}
  A  \; \in \; {{\mathcal{A}}ttributes} ( \Sigma )
\end{equation*}
\begin{thesis}
\end{thesis}
\begin{equation*}
  \exists \; \; \lim_{n \rightarrow \infty} \frac{ N( A | \vec{C}(n) ) }{n}
\end{equation*}

where $N( A | \vec{C}(n) )$ denotes the number of elements of the
prefix $ \vec{C}(n) $ of C of length n  for which the attribute A
holds.

Given an \textbf{attribute} $ A \in  {\mathcal{A}}ttributes (
\Sigma ) $ of a \textbf{collective} $ C \in
{\mathcal{C}}ollectives $ the axiom\ref{ax:axiom of convergence}
make consistent the following definition:
\begin{definition}
\end{definition}
VON MISES' FREQUENTISTIC PROBABILITY OF A IN C:
\begin{equation}\label{eq:von Mises' definition of probability}
  P_{VM} ( A | C ) := lim_{n \rightarrow \infty} \frac{ N( A | \vec{C}(n) ) }{n}
\end{equation}
Let us then introduce the following basic definition:
\begin{definition} \label{def:gambling strategy}
\end{definition}
GAMBLING STRATEGY:

$ S : \Sigma^{\star} \stackrel {\circ}{\rightarrow} \{ 0 , 1 \} $

\smallskip

where, following the notation of \cite{Odifreddi-89}, $ f : A
\stackrel {\circ}{\rightarrow}  B $ denotes a \textit{partial
function} from A to B, i.e. a total function $ f: HALTING(f)
\rightarrow B $, with $  HALTING(f) \subseteq A $ called the
\textit{halting set of f}. If $ x \in A - HALTING(f) $ we will
say that \textit{f doesn't halt on the input x} and will denote
it by $ f(x) = \uparrow $.

Given a gambling strategy S:
\begin{definition} \label{def:subsequence extraction function}
\end{definition}
SUBSEQUENCE EXTRACTION FUNCTION INDUCED BY S:

$ EXT[S] : \Sigma^{\infty} \; \rightarrow \; \Sigma^{\infty} $ :

\begin{equation}
   EXT[S] (  x_{1}  x_{2}  \cdots  ) \; := \; \text{ordered concatenation}
   ( \{ x_{n} \, :
    \,  S ( x_{1}  \cdots x_{n-1} ) = 1 , n \in {\mathbb{N}}_{+} \} )
\end{equation}
The name in the definition\ref{def:subsequence extraction
function} is justified by the fact that obviously:
\begin{equation}
  EXT[S] ( \bar{x} ) \leq_{s}  \bar{x} \; \; \forall \bar{x}
  \in \Sigma^{\infty}
\end{equation}
where $  \leq_{s} $ is the following:
\begin{definition}
\end{definition}
SUBSEQUENCE ORDERING RELATION ON $ \Sigma^{\infty} $
\begin{equation}
  \bar{x} \leq_{s} \bar{y} := \text{$ \bar{x} $ is a subsequence of $ \bar{y} $ }
\end{equation}
\begin{example} \label{ex:bet on the last result}
\end{example}
BET EACH TIME ON THE LAST RESULT

Considered the binary alphabet $ \Sigma \; := \; \{ 0 ,1 \} $, let
us analyze the following gambling strategy:
\begin{equation}
  S ( x_{1}  \cdots  x_{n} ) \; := \;
  \begin{cases}
    \uparrow  &   \text{if $ n = 0 $}, \\
     x_{n}  &   \text{otherwise}
  \end{cases} \; \;  x_{1}  \cdots  x_{n} \in \Sigma^{n} , n \in \mathbb{N}
\end{equation}
and the \textit{subsequence extraction function} $EXT[S]$ it
gives rise to.

Clearly we have that:

\begin{tabular}{|c|c|}
  $\vec{x}$ & $S( \vec{x} )$ \\ \hline
  $\lambda$ & $\uparrow$ \\
  0 & 0 \\
  1 & 1 \\
  00 & 0 \\
  01 & 1 \\
  10 & 0 \\
  11 & 1 \\
  000 & 0 \\
  001 & 1 \\
  010 & 0 \\
  011 & 1 \\
  100 & 0 \\
  101 & 1 \\
  110 & 0 \\
  111 & 1 \\
  0000 & 0 \\
  0001 & 1 \\
  0010 & 0 \\
  0011 & 1 \\
  0100 & 0 \\
  0101 & 1 \\
  0110 & 0 \\
  0111 & 1 \\
  1000 & 0 \\
  1001 & 1 \\
  1010 & 0 \\
  1011 & 1 \\
  1100 & 0 \\
  1101 & 1 \\
  1110 & 0 \\
  1111 & 1 \\ \hline
\end{tabular}

\smallskip

Furthermore we have, clearly, that:
\begin{align*}
  EXT[S] ( 0^{\infty} ) \; & = \; \lambda \\
  EXT[S] ( 0^{\infty}) \; & = \;  1^{\infty} \\
  EXT[S] ( 01^{\infty}\cdots ) \; & = \; 0^{\infty} \\
  EXT[S] ( 10^{\infty} ) \; & = \; 0^{\infty} \\
  EXT[S] ( \bar{x}_{Champernowne} ) \; & = \; 0101\cdots
\end{align*}
where $ \bar{x}_{Champernowne} $ is the Champernowne sequence
defined as the lexicografic ordered concatenation of the binary
strings:
\begin{equation*}
  \bar{x}_{Champernowne} \; = \; 0100011011000001010011100101110111 \cdots
\end{equation*}

\begin{example} \label{eq:bet on the less frequent letter}
\end{example}
BET ON THE LESS FREQUENT LETTER

Considered again the binary alphabet $ \Sigma \; := \; \{ 0 ,1 \}
$, let us analyze the following gambling strategy:
\begin{equation}
  S ( \vec{x} ) \; = \;
  \begin{cases}
    \uparrow & \text{if $ \vec{x} = \lambda $ or $ N_{0} ( \vec{x} ) = N_{1} ( \vec{x} )$} , \\
    1 & \text{if  $N_{0} ( \vec{x} ) > N_{1} ( \vec{x} )$} , \\
    0 & \text{otherwise}.
   \end{cases}
\end{equation}
where $ N_{0} ( \vec{x} ) , N_{1} ( \vec{x} ) $ denote the number
of, respectively, zeros and ones in the string $ \vec{x} $.

We have that:

\smallskip

\begin{tabular}{|c|c|}
  $\vec{x}$ & $S( \vec{x} )$ \\ \hline
  $\lambda$ & $\uparrow$ \\
  0 & 1 \\
  1 & 0 \\
  00 & 1   \\
  01 & $\uparrow$ \\
  10 & $\uparrow$ \\
  11 & 0 \\
  000 & 1 \\
  001 & 1  \\
  010 & 1 \\
  011 & 0 \\
  100 & 1 \\
  101 & 0 \\
  110 & 0 \\
  111 &  0 \\
  0000 & 1 \\
  0001 & 0 \\
  0010 & 1 \\
  0011 & $\uparrow$ \\
  0100 & 1 \\
  0101 & $\uparrow$ \\
  0110 & $\uparrow$ \\
  0111 & 1 \\
  1000 & 1 \\
  1001 & $\uparrow$ \\
  1010 & $\uparrow$ \\
  1011 & 0 \\
  1100 & $\uparrow$ \\
  1101 & 0 \\
  1110 & 0 \\
  1111 & 0 \\ \hline
\end{tabular}

\smallskip

As to the extraction function of S:
\begin{align*}
  EXT[S] ( 0^{\infty} ) \; & = \; 0^{\infty} \\
  EXT[S] ( 1^{\infty} ) \; & = \;  \lambda \\
  EXT[S] ( 01^{\infty} ) \; & = \; 1^{\infty} \\
  EXT[S] ( 10^{\infty} ) \; & = \; \lambda \\
  EXT[S] ( \bar{x}_{Champernowne} ) \; & = \; 10011011\cdots
\end{align*}

Denoted by $ {\mathcal{S}}trategies ( {\mathcal{C}}ollectives ) $
the \textbf{set of gambling strategies} concerning $
{\mathcal{C}}ollectives $, we can formalize the \textbf{Law of
Excluded Gambling Strategies} by the following:
\begin{axiom} \label{ax:axiom of randomness}
\end{axiom}
AXIOM OF RANDOMNESS
\begin{hypothesis}
\end{hypothesis}
\begin{equation*}
  S \, \in \,  {{\mathcal{S}}trategies}_{admissible} ( {\mathcal{C}}ollectives )
\end{equation*}
\begin{equation*}
  C \, \in \, {\mathcal{C}}ollectives
\end{equation*}
\begin{equation*}
  A  \, \in \, {{\mathcal{A}}ttributes} ( \Sigma )
\end{equation*}
\begin{thesis}
\end{thesis}
\begin{equation*}
  P_{VM}( \,A  \,| \, EXT[S] (C) \,  ) \; = \;  P_{VM} ( A | C )
\end{equation*}
where $ {{\mathcal{S}}trategies}_{admissible} (
{\mathcal{C}}ollectives ) \; \subseteq \; {\mathcal{S}}trategies (
{\mathcal{C}}ollectives )  $ is the \textbf{set of admissible
gambling strategies} whose mathematical characterization will
lead us, in the next sections, to the heart of Classical
Algorithmic Information Theory.
\newpage
\section{Classical Gambling in the framework of Classical Statistical Decision
Theory} \label{sec:Classical Gambling in the framework of
Classical Statistical Decision Theory}

Classical Statistical Decision Theory \cite{French-Rios-Insua-00}
concerns the following situation:

a \textit{decision maker} have to make a  single action $ a \in
{\mathcal{A}}ctions $ from a space $ {\mathcal{A}}ctions $ of
possible actions.

Features that are unknown about the external world are modelled
by an unknown state of nature $ s \in {\mathcal{S}}tates $ in a
set $ {\mathcal{S}}tates $ of possible states of nature.

The consequence $ c ( a , s ) \in {\mathcal{C}}onsequences $ of
his choice depends both on the action chosen and on the unknwown
state of nature.

Before making his decision the decision maker may observe an
outcome $ X = x $ of an experiment, which depends on the unknown
state s. Specifically the observation X is drawn from a
distribution $ P_{X} ( \cdot | s ) $.

His objectives are encoded in a real valued \textit{utility
function} $ u ( a , s ) $.

Let us assume that the \textit{decision maker} knows the
\textit{action space} $ {\mathcal{A}}ctions $, \textit{state
space} $ {\mathcal{S}}tates $ and \textit{consequence space} $
{\mathcal{C}}onsequences $, along with the probability
distribution and the \textit{utility function}.

His problem is:

\textbf{observe $ X = x $ and then choose an action $ d(x) \in
{\mathcal{A}}ctions $, using the information that $ X = x $, to
maximize, in some sense, $ u ( d ( x ) , s ) $ }.

Every decision process may obviously  be seen as a gambling
situation: the \textit{action space} $ {\mathcal{A}}ctions$ may
be seen as the set of possible bets of the decision maker, that
we will call from here and beyond the \textit{gambler}, while the
\textit{utility function} gives the \textit{payoff}.

Let us consider, in particular, the following gambling situation:

in the city's \textit{Casino} at each turn $ n \in \mathbb{N} $
the croupier tosses a fair coin.

Before the $ n^{th} $ toss the gambler can choose among one of the
possbile choices:
\begin{itemize}
  \item to bet one fiche on \textit{head}
  \item to bet one fiche on \textit{tail}
  \item not to play at that turn
\end{itemize}
Leaving all the philosophy behind its original foundational
purpose we can, now, from inside the standard Kolomogorovian
measure-theoretic formalization of Classical Probability Theory,
appreciate the very intuitive meaning lying behind Von Mises'
axioms.

Let us indicate  by $ X_{n} $ the random variable on the binary
alphabet $ \Sigma := \{ 0 , 1 \} $ (where we will assume from
here and beyond, that $ head = 1 $ and $ tail = 0 $)
corresponding to the $ n^{th} $ coin toss and by $ x_{n} \in
\Sigma $ the result of the $ n^{th} $ coin toss.

Let us, furthermore, denote by $ \bar{x} \; := \; (  x_{1} ,
x_{2} , , \cdots )\in \Sigma^{\infty} $ the sequence of all the
results of the coin tosses and by $ \vec{x}(n) \in \Sigma^{n} $
its $ n^{th} $ prefix.

By hypothesis $\{ X_{n} \}_{n \in \mathbb{N}} $ is a Bernoulli($
\frac{1}{2}$)  discrete-time stochastic process over $ \Sigma $.

A gambling strategy $ S : \Sigma^{\star} \stackrel
{\circ}{\rightarrow} \{ 0 , 1 \} $ determines the gambler's
decision at the $ n^{th} $ turn in the following way:
\begin{itemize}
  \item if $ S( \vec{x}(n-1) ) \; = \; 1 $ he bets on \textit{head}
  \item if $ S( \vec{x}(n-1) ) \; = \; 0 $ he bets on \textit{tail}
  \item if $  S( \vec{x}(n-1) ) \; = \; \uparrow $ he doesn't bet at that turn
\end{itemize}

The situation may be simulated by the following Mathematica code:
\newpage

\includegraphics[scale=0.85]{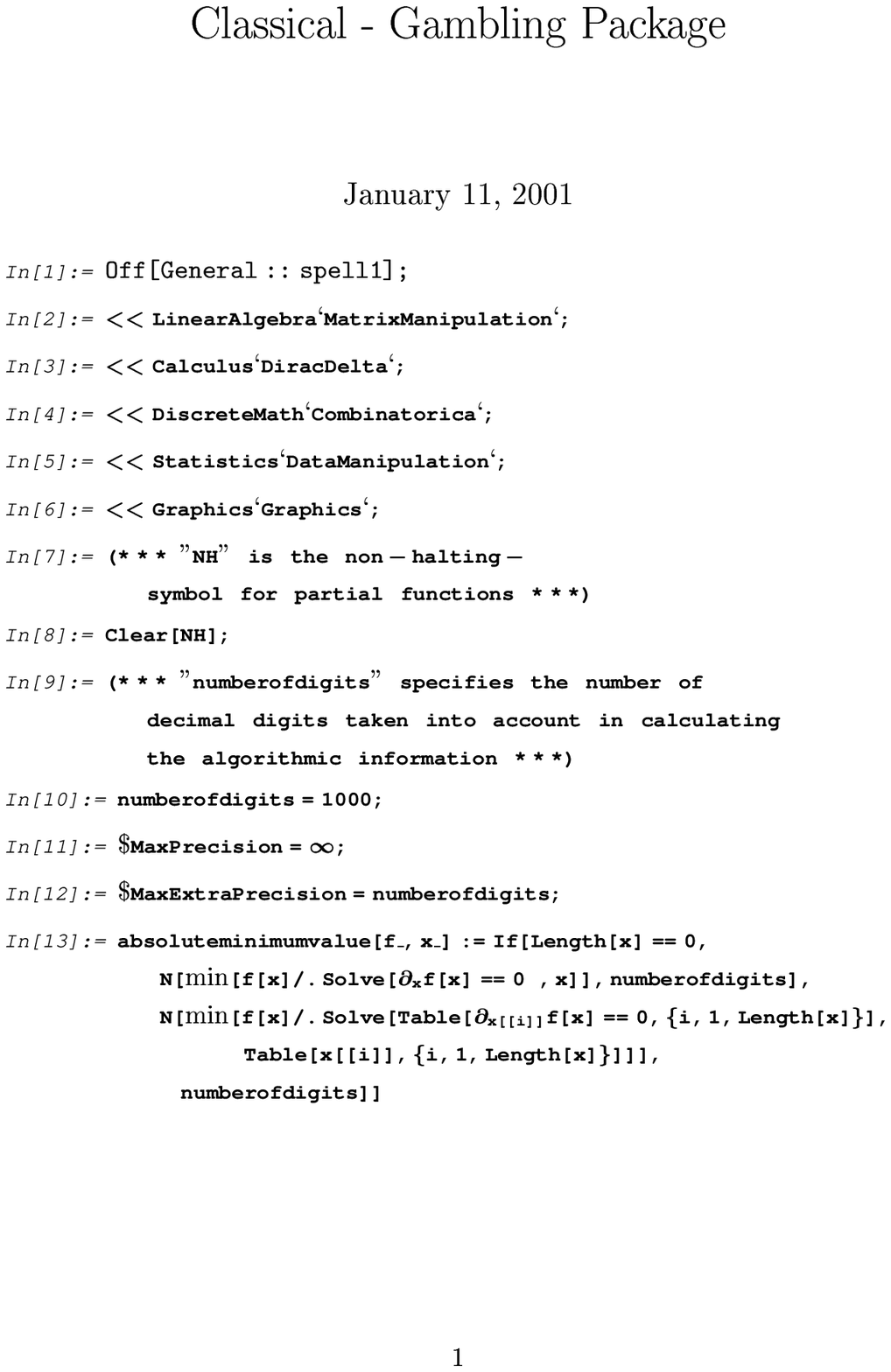}
\includegraphics[scale=0.85]{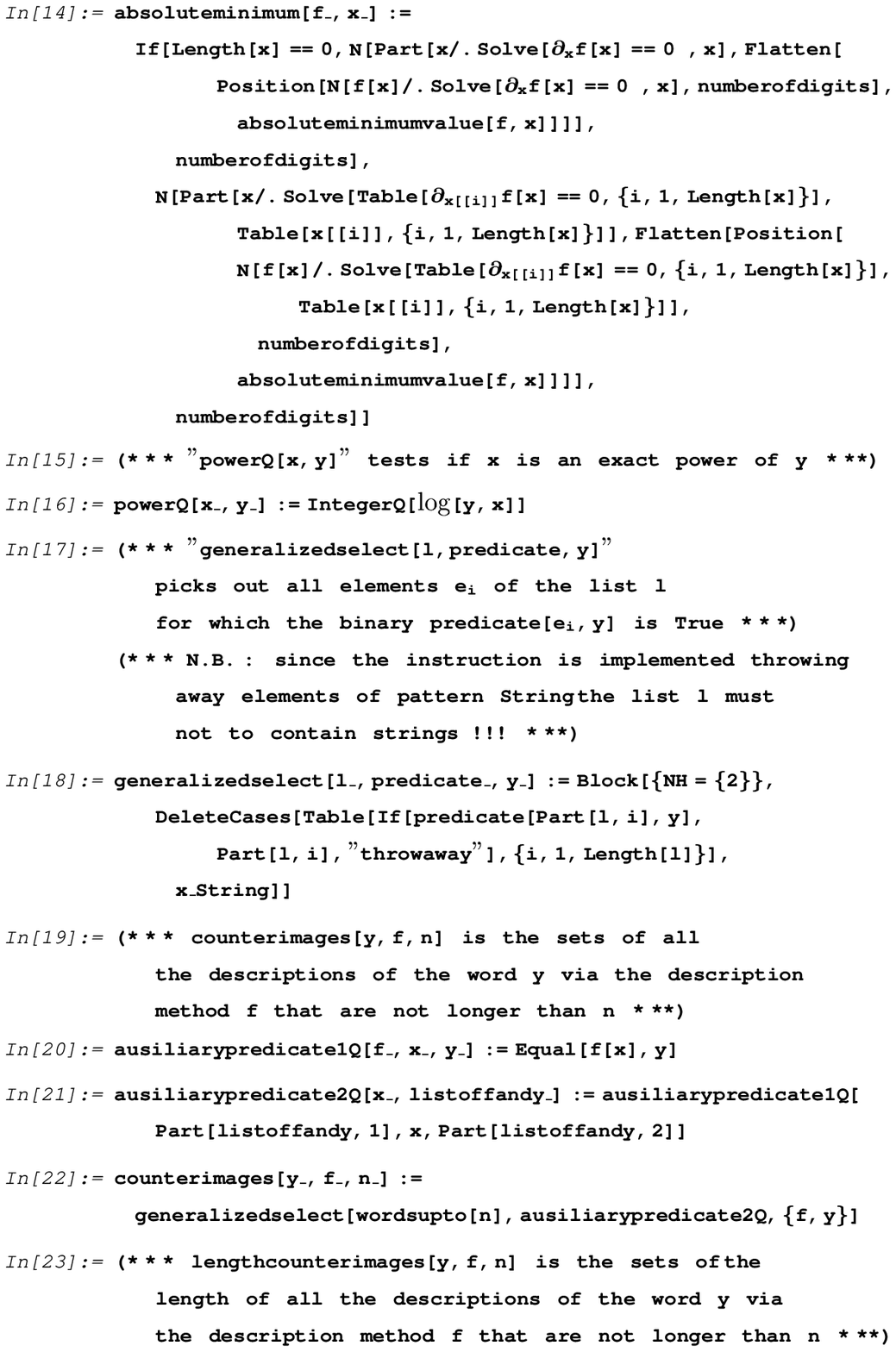}
\includegraphics[scale=0.85]{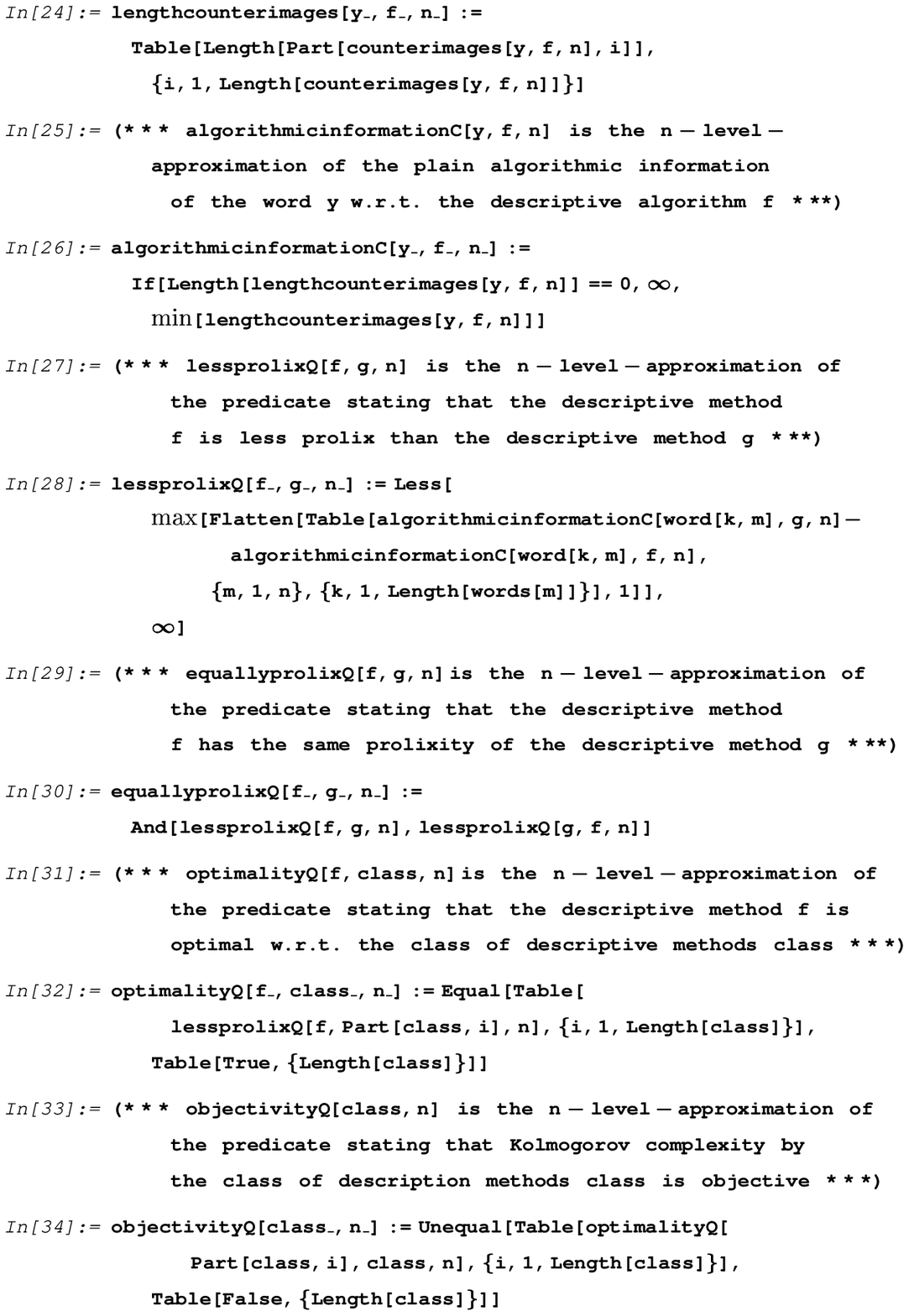}
\includegraphics[scale=0.85]{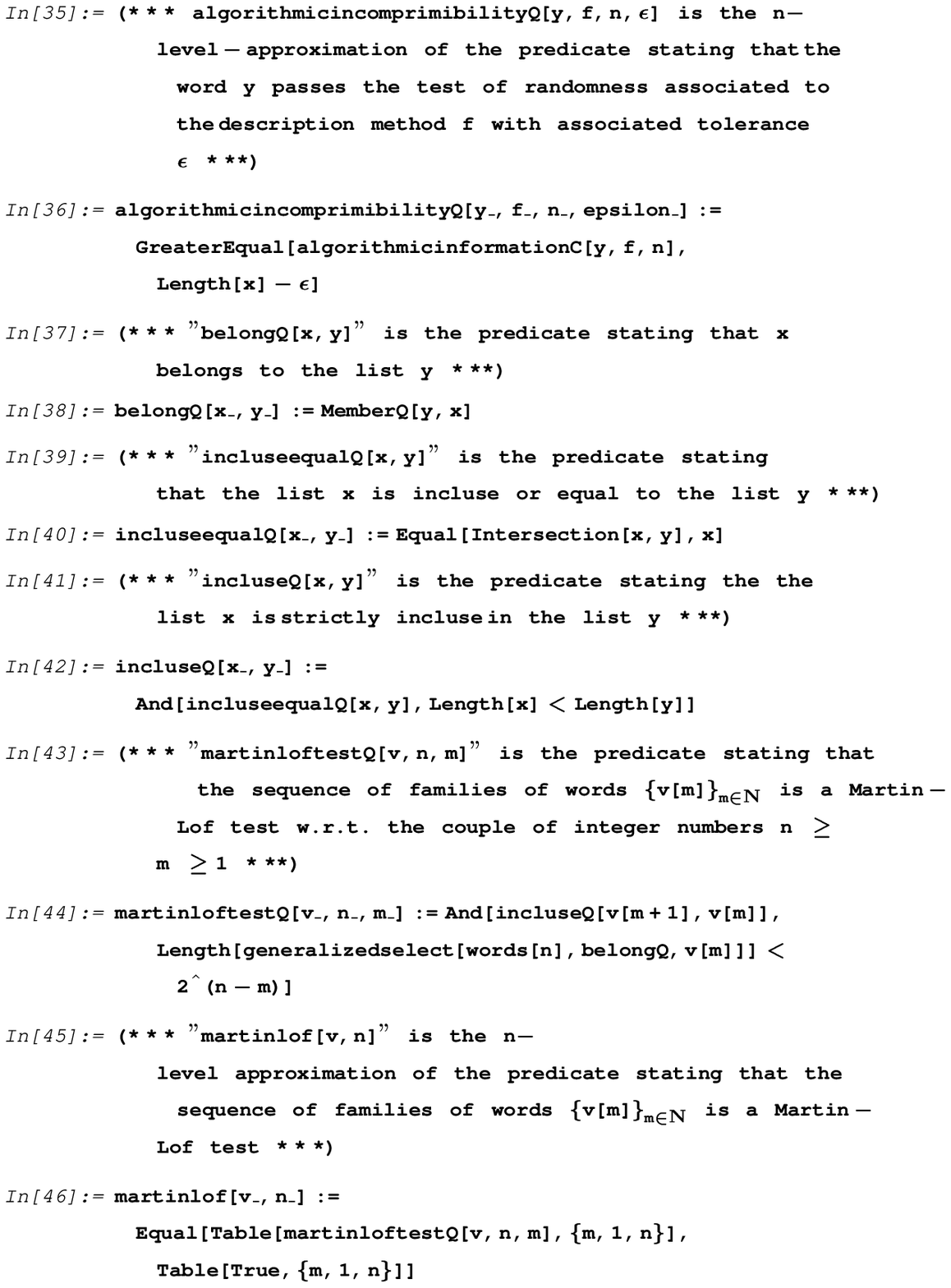}
\includegraphics[scale=0.85]{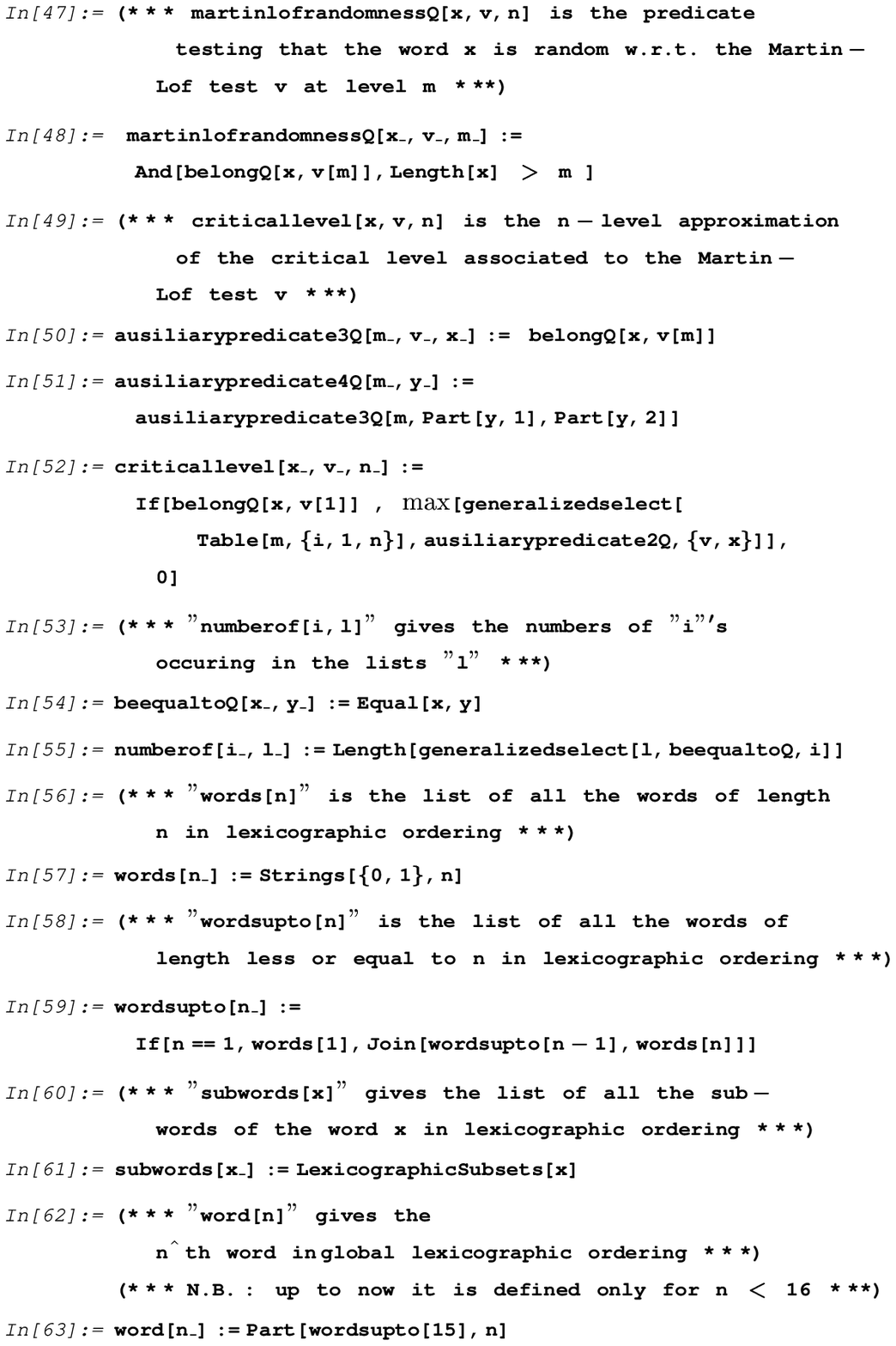}
\includegraphics[scale=0.85]{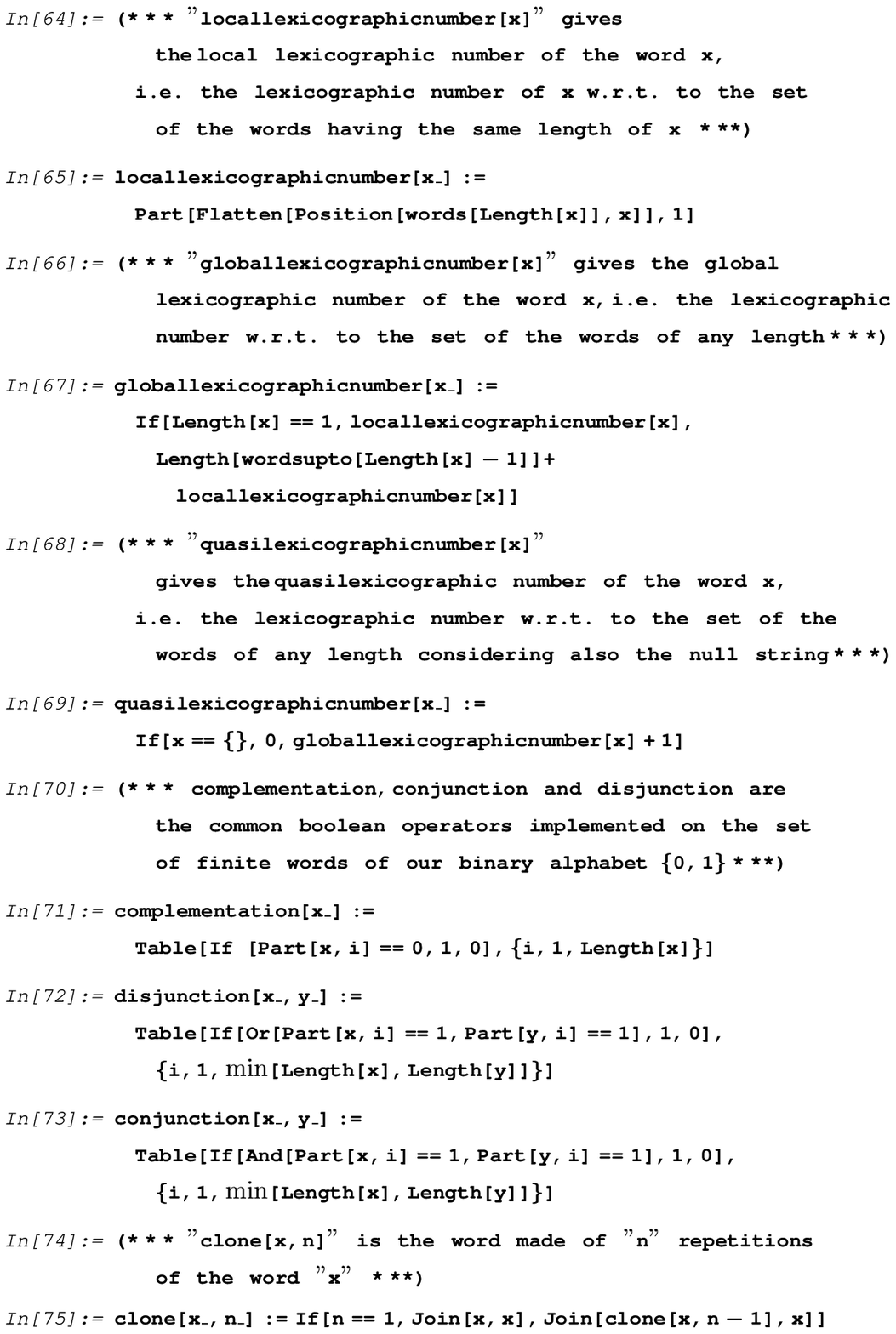}
\includegraphics[scale=0.85]{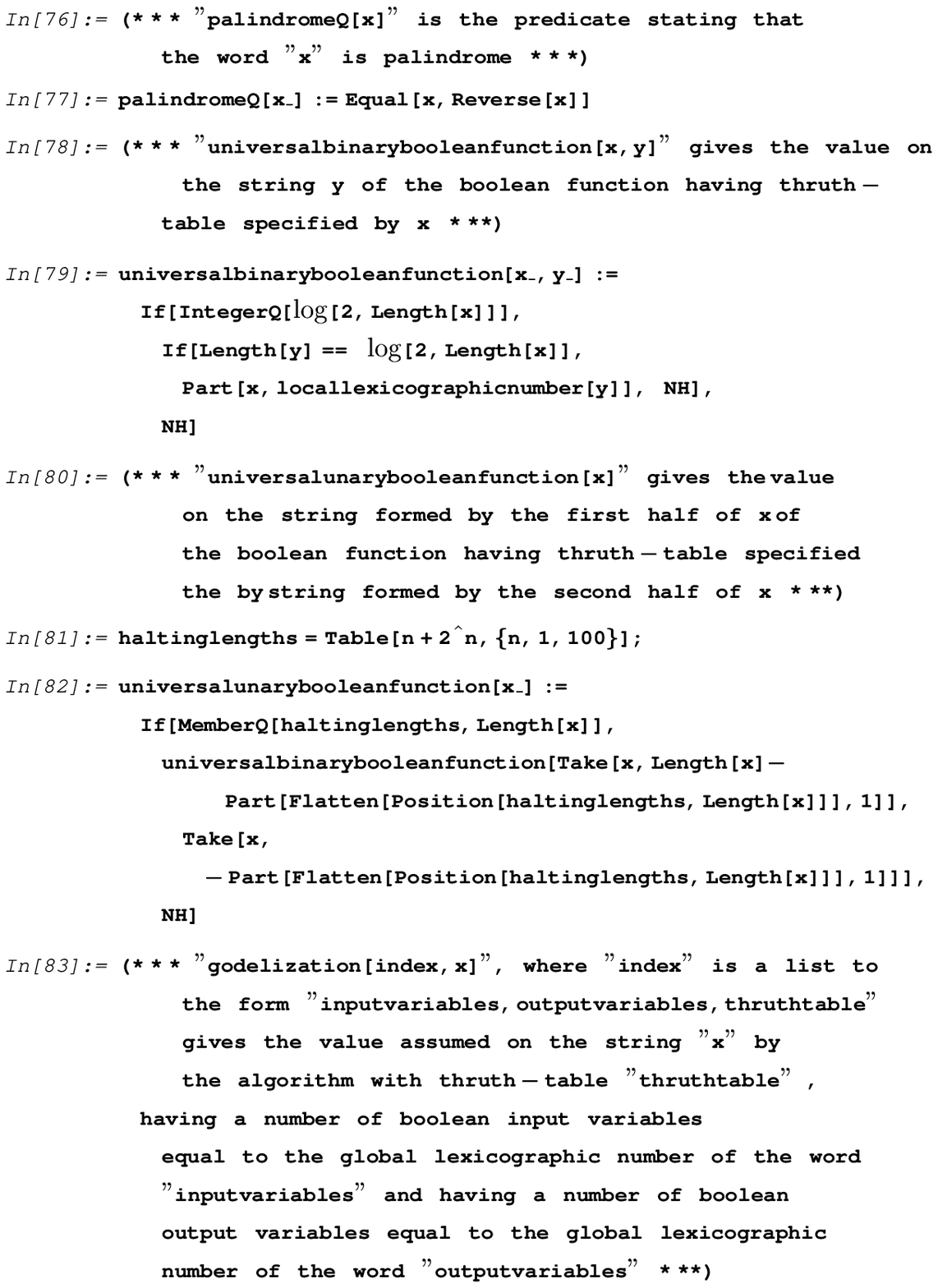}
\includegraphics[scale=0.85]{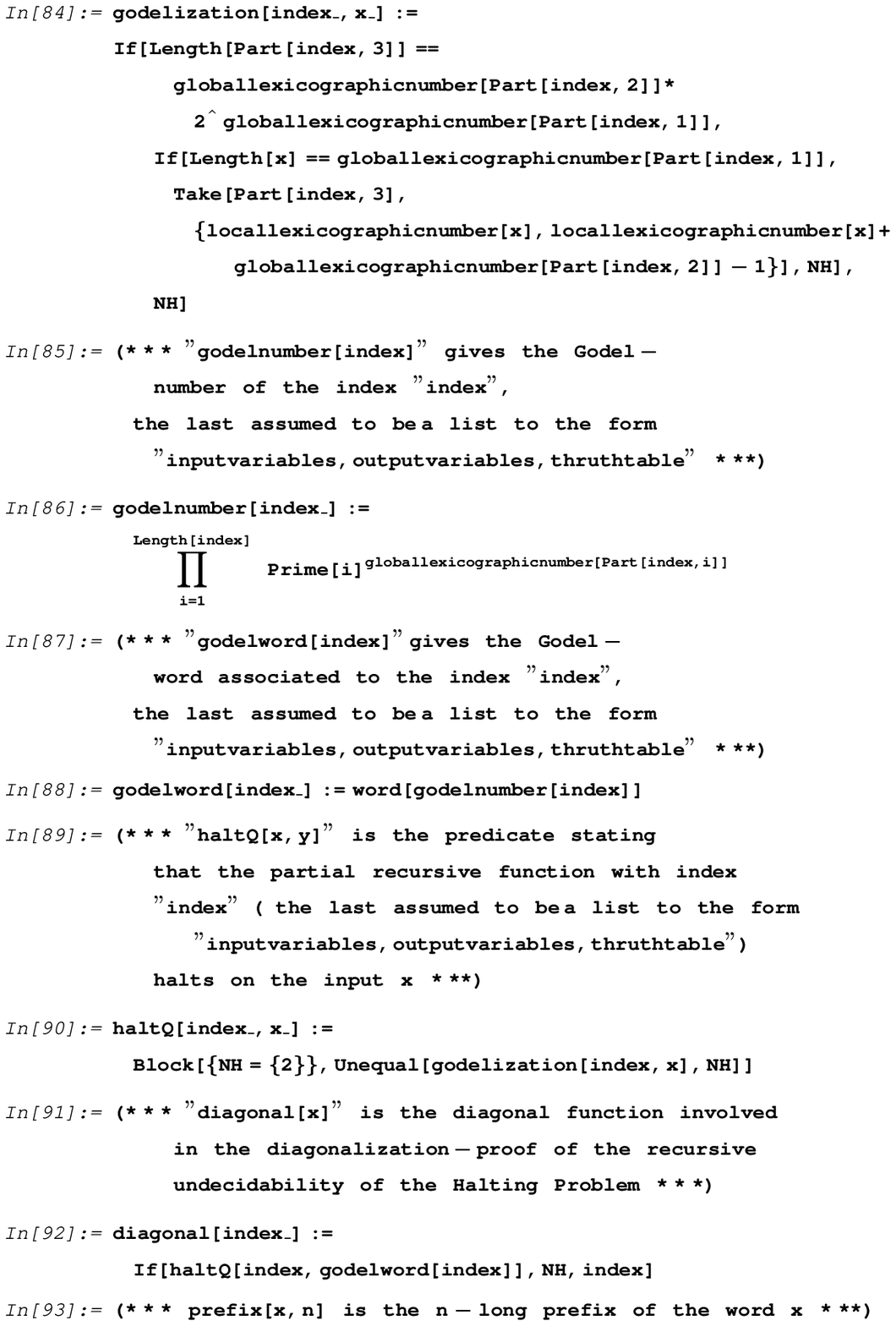}
\includegraphics[scale=0.85]{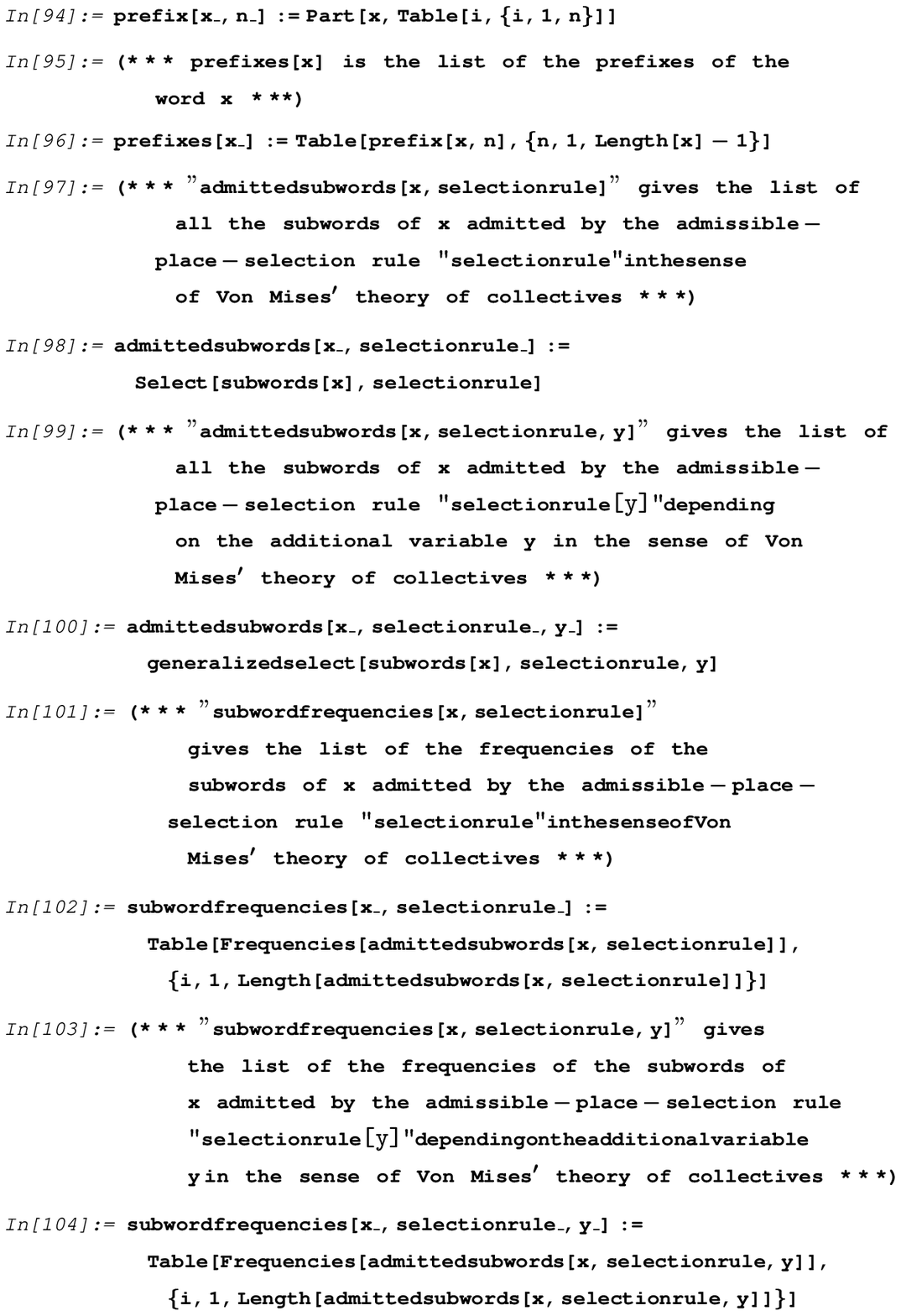}
\includegraphics[scale=0.85]{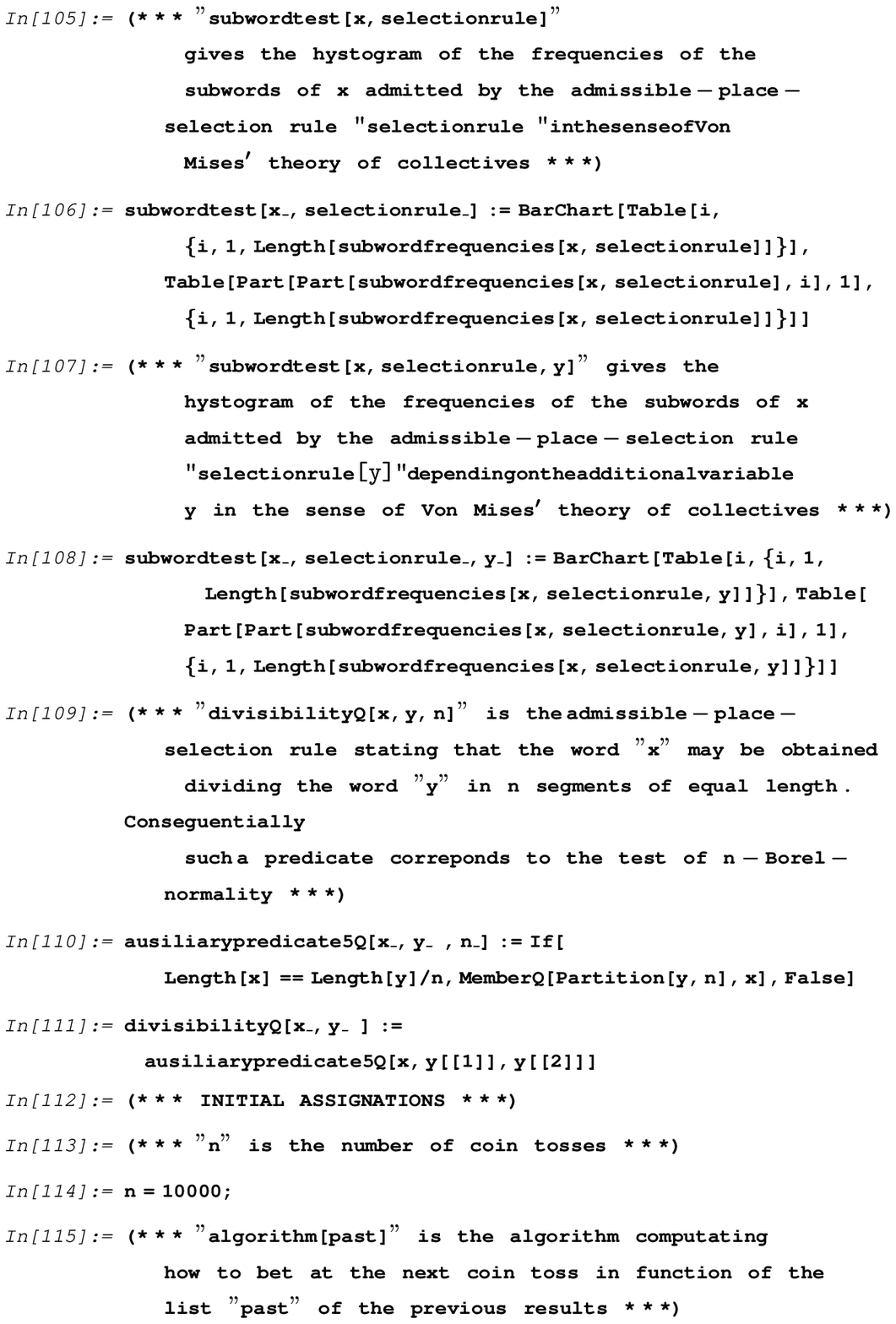}
\includegraphics[scale=0.85]{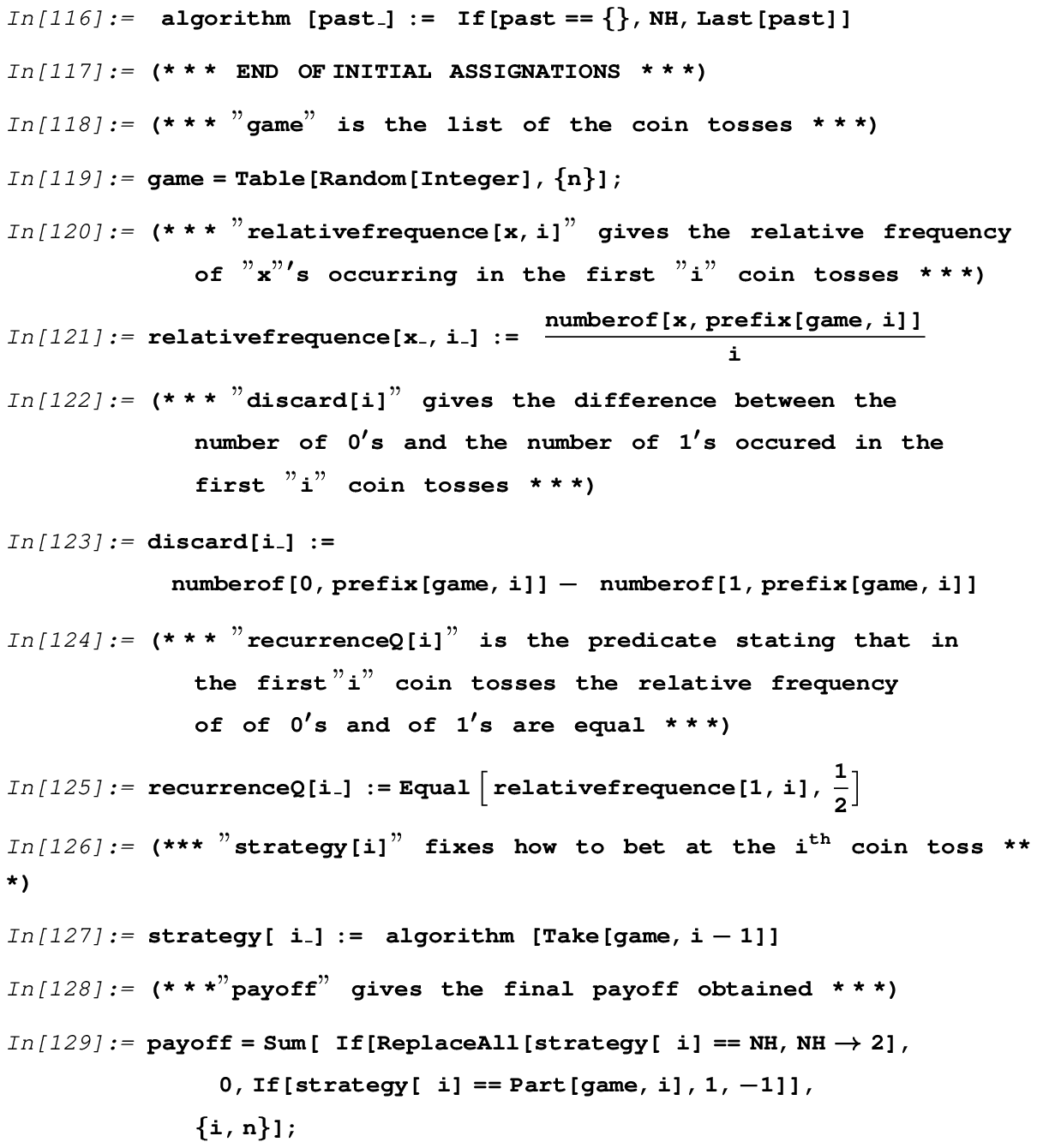}
\newpage
where the initial assignations may be arbitarily variated from
their default: a number of $ 10000 $ coin tosses and the adoption
of the gambling strategy discussed in the following example

\begin{example}
\end{example}

APPLYING TO THE CASINO THE GAMBLING STRATEGY OF
EXAMPLE\ref{ex:bet on the last result}

Let us suppose that the first 10 coin tosses give the  following
string of results: $ \vec{x}(n) \; = \; 1101001001 $

Our evening to Casino may be told by the following table:

\smallskip

\begin{tabular}{|c|c|c|c|}
TOSS &  RESULT OF THE TOSS & BET MADE ABOUT THAT TOSS &  PAYOFF  \\
\hline
  1  &          1          &         no bet           &        0         \\
  2  &          1          &          1               &       +1        \\
  3  &          0          &          1               &        0       \\
  4  &          1          &          0               &       -1        \\
  5  &          0          &          1               &       -2        \\
  6  &          0          &          0               &       -1        \\
  7  &          1          &          0               &       -2        \\
  8  &          0          &          1               &       -3        \\
  9  &          0          &          0               &       -2        \\
  10 &          1          &          0               &       -3        \\ \hline
\end{tabular}

\smallskip

As we see $ PAYOFF(10) \; = \; -3 $.

\begin{example}
\end{example}
APPLYING TO THE CASINO THE GAMBLING STRATEGY OF
EXAMPLE\ref{eq:bet on the less frequent letter}

\smallskip

Also this gambling situation may of course be simulated with the
previously introduced Mathematica code by, simply,  changing the
initial assignation of the function \textit{algorithm[past]} in
the following way:

\begin{multline}
  algorithm[past_{-} ] \; := \; If[ numberof[1,past] ==
  numberof[0,past] , NH , \\ If [ numberof[1,past] >
  numberof[0,past] , 0 , 1 ]]
\end{multline}

\smallskip

Let us suppose again that the first 10 coin tosses give the
following string of results: $ \vec{x}(n) \; = \; 1101001001 $.

Our evening to Casino may be told by the following table:

\smallskip

\begin{tabular}{|c|c|c|c|}
TOSS &  RESULT OF THE TOSS & BET MADE ABOUT THAT TOSS & PAYOFF
\\ \hline
  1  &          1          &         no bet           &       0          \\
  2  &          1          &           0              &      -1          \\
  3  &          0          &           0              &       0          \\
  4  &          1          &           0              &      -1          \\
  5  &          0          &           0              &       0          \\
  6  &          0          &           0              &      +1          \\
  7  &          1          &         no bet           &      +1           \\
  8  &          0          &           0              &      +2          \\
  9  &          0          &         no bet           &      +2          \\
  10 &          1          &           1              &      +3          \\ \hline
\end{tabular}

\smallskip

As we see $ PAYOFF(10) \; = \; +3  $.

\bigskip

The probability distribution of the string $ \vec{x}_{n} $ is the
uniform distribution on $ \Sigma^{n} $ :
\begin{equation}
  Prob [ \vec{x}(n) = \vec{y} ] \; = \; \frac{1}{2^{n}} \; \; \forall
  \vec{y} \in \Sigma^{n} , \forall n \in \mathbb{N}
\end{equation}

When $ n \rightarrow \infty $ such a distribution tends to the
following:
\begin{definition}
\end{definition}
UNBIASED PROBABILITY MEASURE ON $ \Sigma^{\infty} $:

$ P_{unbiased} \, : \, 2^{ \Sigma^{\infty}} \; \stackrel{\circ
}{\rightarrow} \; [0,1]$ :

\begin{align}
  HALTING(P_{unbiased}) \; & = \; {\mathcal{F}}_{cylinder} \\
  P_{unbiased} ( \Gamma_{\vec{x}} ) \; & \equiv \;  \frac{1}{2^{| \vec{x} |}} \; \;
  \forall \, \vec{x} \, \in \, \Sigma^{\star}
\end{align}
with $ | \vec{x} | $ denoting the length of the string $ \vec{x}
$ and where:
\begin{definition}
\end{definition}
CYLINDER SET W.R.T. $ \vec{x} \, =  ( x_{1} , \ldots , x_{n} ) \,
\in \, \Sigma^{\star} $:
\begin{equation} \label{eq:cylinder set}
\Gamma_{\vec{x}} \; \equiv \; \{ \bar{y} = ( y_{1} , y_{2} ,
\ldots ) \in \Sigma^{\infty} \; : \; y_{1} = x_{1} , \ldots ,
y_{n} = x_{n} \}
\end{equation}
\begin{definition}
\end{definition}
CYLINDER - $ \sigma $ - ALGEBRA ON $ \Sigma^{\infty} $:
\begin{equation}
  {\mathcal{F}}_{cylinder} \; \equiv \;  \sigma- \text{algebra generated by}  \{ \Gamma_{\vec{x}} \, : \, \vec{x} \in \Sigma^{\star}
  \}
\end{equation}

Clearly the possible \textbf{attributes} of a letter on the binary
alphabet are :
\begin{itemize}
  \item $ a_{1} \; := \;   << \text{to be 1} >> $
  \item $ a_{0} \; := \;  << \text{to be 0} >> $
\end{itemize}
so that:
\begin{equation}
  {{\mathcal{A}}ttributes} ( \Sigma ) \; = \; \{ a_{1} , a_{0} \}
\end{equation}
Whichever $ {\mathcal{C}}ollectives\; \subset \;  \Sigma^{\infty}
$ is the axiom\ref{ax:axiom of convergence} is, from inside the
standard kolmogorovian measure-theoretic foundation, an immediate
corollary of the Law of Large Numbers.

As far as axiom\ref{ax:axiom of randomness} is concerned, anyway,
the situation is extraordinarily subtler.

Every \textbf{intrinsic regularity} of $ \vec{x}(n) $ could have
been encoded by the gambler in a proper \textbf{winning strategy
up to the $ n^{th} $ turn}.

The same definition of what a \textbf{winning strategy} is
requires some caution:

we can, indeed, give two possible definitions of such a concept:
\begin{definition}[AVERAGE-WINNING STRATEGY UP TO THE $ n^{th} $
TOSS] \label{def:average winning strategy}

a strategy so that the expectation value of the payoff after the
first n tosses \textbf{payoff(n) } is greater than zero
\end{definition}

The fact the a strategy is average-winning doesn't imply that the
payoff after the $ n^{th} $ toss will be strictly positive with
certainty: it happens if we are lucky.

Let us now introduce a weaker notion of a winning strategy:

\begin{definition}[LUCKY-WINNING STRATEGY UP TO THE $ n^{th} $
TOSS] \label{def:lucky winning strategy}

a strategy so that the the probability that the  payoff after the
first n tosses \textbf{payoff(n) } is greater than zero is itself
greater than zero
\end{definition}

For finite n every strategy is obviously lucky-winning.

Let us now consider the limit $ n \rightarrow \infty $.

By purely measure-theoretic considerations we may easily prove the
following:
\begin{theorem} \label{th:weak law of excluded gambling systems}
WEAK LAW OF EXCLUDED GAMBLING STRATEGIES

For $ n \rightarrow \infty $ the set of  the
\textit{average-winning strategies} tends to the null set
\end{theorem}
\begin{proof}
Given a gambling strategy $ S : \Sigma^{\star} \stackrel
{\circ}{\rightarrow} \{ 0 , 1 \} $ we have clearly that:
\begin{multline} \label{eq:conditional payoff at step n}
  E[payoff(n) | payoff(n-1) ] \; = \; payoff(n-1) + \\
  If[ S ( \vec{x}_{n-1} ) = \uparrow , 0 ,  \frac{1}{2} \\
If[ S ( \vec{x}_{n-1} )= 1 , 1 , -1 ] + \frac{1}{2} \\
   If[ S ( \vec{x}_{n-1} ) = 0 , 1 , -1]] \; = \; payoff(n-1)  \; \; \forall n \in \mathbb{N}
\end{multline}
(where I have adopted Mc Carthy's LISP conditional notation
\cite{Mc-Carthy-60} popularized by Wolfram's \textit{Mathematica}
\cite{Wolfram-96}).

Furhermore:
\begin{equation} \label{global payoff at step n}
    E[payoff(n)] \; = \; \sum_{k = -n+1}^{n-1} P[payoff(n-1) = k] \,  E[payoff(n) | payoff(n-1)
    ] \; \; \forall n \in \mathbb{N}
\end{equation}

We will prove that $ \lim_{n \rightarrow \infty} E[ payoff(n) ]
\; = \; 0 $ by proving by induction on n that $ E[ payoff(n) ] \;
= \; 0 \; \; \forall n \in {\mathbb{N}} $.

That $ E[payoff(1)] \; = \; 0 $ follows immediately by the fact
that $ S ( \lambda ) \; = \; \uparrow \; \; \forall S $.

We have, conseguentially, simply to prove that $  E[payoff(n-1)]
\; = \; 0 \; \; \Rightarrow \; \; E[payoff(n)] \; = \; 0 \; \;
\forall S $.

This is, anyway, an obvious conseguence of the equations
eq.\ref{eq:conditional payoff at step n} and eq.\ref{global payoff
at step n}
\end{proof}

\bigskip

Theorem\ref{th:weak law of excluded gambling systems} is not,
anyway, a great assurance for Casino's owner:

in fact it doesn't exclude that the gambler, if  enough lucky, may
happen to get a positive payoff for $ n \rightarrow \infty $.

What will definitely assure him is the following:
\begin{theorem} \label{strong law of excluded gambling strategies}
STRONG LAW OF EXCLUDED GAMBLING STRATEGIES

For $ n \rightarrow \infty $ the set of  the \textit{lucky-winning
strategies} tends to the null set
\end{theorem}
And here comes the astonishing fact: Theorem\ref{strong law of
excluded gambling strategies} can't be proved with purely
measure-theoretic concepts.

Our approach will consist in taking von Mises' axiom\ref{ax:axiom
of randomness} as a definition of the set of subsequences   to
which such an axiom applies.

Let us then define the \textbf{set of collectives} $
{\mathcal{C}}ollectives\; \subset \; \Sigma^{\infty} $ as the set
of sequences having not enough intrinsic regularity to allow, if
they occur, a lucky-winning strategy. Clearly such a definition
depends on the class $ {{\mathcal{S}}trategies}_{admissible} (
{\mathcal{C}}ollectives )$ of admissible gambling strategies.

It would appear natural ,at first, to admit every gambling
strategy.

But such a choice would lead immediately to conclude that $
{\mathcal{C}}ollectives \; = \; \emptyset $ since given two
gambling strategies $ S_{0} $ and $ S_{1} $  so that:
\begin{equation}
  EXT[S_{i}] ( \bar{x} ) \; \text{ is made only of i } \; i=0,1 \;
  \forall \bar{x} \in \Sigma^{\infty}
\end{equation}
we would have clearly that
\begin{equation}
   P_{VM}( a_{i}  \,| \, EXT[S_{1}] (\bar{x}) \, ) \; \neq P_{VM}( a_{i}  \,| \, EXT[S_{2}]
   (\bar{x}) \; \; \forall \bar{x} \in \Sigma^{\infty}
\end{equation}
The history of the attempts of characterizing in a proper way the
class of the admissible gambling strategies is  very long and
curious \cite{Li-Vitanyi-97}, \cite{Gillies-00} and involved many
people: Church, Copeland, D\"{orge}, Feller, Kamke, Popper,
Reichenbach, Tornier, Waismann and Wald; I will report here only
the conceptually more important contributions:

in the thirties Abraham Wald showed that:
\begin{equation}
   ( cardinality (  {{\mathcal{S}}trategies}_{admissible} ( {\mathcal{C}}ollectives
  ) ) \, = \, \aleph_{0} ) \; \Rightarrow \;( {\mathcal{C}}ollectives
  \neq \emptyset )
\end{equation}

In the fourties,  basing on the observation that gambling
strategies must be effectively followed, Alonzo Church  proposed,
according to his Church Thesis \cite{Odifreddi-89}, to consider
admissible a gambling strategy if and only if it is a
\textbf{partial recursive function}.

With such an assumption:
\begin{equation} \label{eq:Church admissible gambling strategies}
  {{\mathcal{S}}trategies}_{admissible} ( {\mathcal{C}}ollectives
  ) )  \; := \Delta_{0}^{0} ( \Sigma^{\star} )
\end{equation}
that I will adopt from here and beyond,it can be proved that:
\begin{equation}
  P_{Unbiased} ( ( {\mathcal{C}}ollectives
  ) ) ) \; = \; 1
\end{equation}
immediately implying Theorem\ref{strong law of excluded gambling
strategies}
\newpage
\section{Mises-Wald-Church randomness versus Martin L\"{o}f-Solovay-Chaitin randomness}
The intuitive idea underlying behind the notion of von Mises -
Wald - Church collectives is very similar to the idea underlying
the most fascinating concept of Classical Algorithmic Information
Theory, namely \textbf{classical algorithmic randomness}. For this
reason  we will refer, from here and beyond, to von Mises - Wald -
Church collectives also as  the von Mises - Wald - Church random
sequences and will denote them by $ RANDOM_{MWC} (
\Sigma^{\infty} ) $.

The universally accepted  notion of classical algorithmic
randomness , i.e. Martin L\"{o}f-Solovay-Chaitin randomness, is,
anyway, stronger \cite{Calude-94},\cite{Li-Vitanyi-97}:
\begin{equation}
    RANDOM_{MWC} ( \Sigma^{\infty} ) \; \subset \;  RANDOM_{MLSC} ( \Sigma^{\infty} )
\end{equation}
Surprisingly it has been proved \cite{Svozil-93} even that:
\begin{equation}
  cardinality ( RANDOM_{MLSC} ( \Sigma^{\infty} ) \, - \, RANDOM_{MWC} ( \Sigma^{\infty}
  ) ) \; = \; \aleph_{1}
\end{equation}
We will restrict here to analize Martin L\"{o}f
algorithmic-measure-way of defining $ RANDOM_{MLSC} (
\Sigma^{\infty} ) $.

Given a  classical  probability space $ CPS \; \equiv \; ( \, M
\, ,  \, \mu  \, ) $:
\begin{definition} \label{def:classical null set}
\end{definition}
$ S \; \subset  \; M $  IS A NULL SET OF CPS :
\begin{equation}
  \forall \epsilon > 0 \; \exists \mu_{ \epsilon } \in HALTING(\mu)  \; : \; S
  \subset F_{ \epsilon } \; and \; \mu( F_{ \epsilon } ) < \epsilon
\end{equation}

Let us introduce the following notions:
\begin{definition}
\end{definition}
UNARY PREDICATES ON M :
\begin{equation}
  {\mathcal{P}} ( M ) \; \equiv \; \{ p( x ) \, : \, \text{ predicate about } x \in M  \}
\end{equation}
\begin{definition}
\end{definition}
TYPICAL PROPERTIES OF CPS:
\begin{equation}
 {\mathcal{P}}  (  CPS  )_{TYPICAL} \; \equiv \; \{ \, p ( x )  \in  {\mathcal{P}} ( M ) \, :
   \{ x \in M \, : \, p ( x ) \text{ doesn't hold } \}
   \; \text{is a null set} \}
\end{equation}
\begin{example}
\end{example}
TYPICAL PROPERTIES OF A DISCRETE CLASSICAL PROBABILITY SPACE

If CPS is discrete-finite $ ( \, M \; = \; \{ a_{1} \, , \ldots \,
a_{n} \} \, ) $  or  discrete-infinite  $ ( \, M \; = \; \{ a_{n}
\}_{n \in \mathbb{N}} \, ) $ it is natural to assume that $ \mu (
\{ a_{i} \} ) > 0  \; \forall i $  since an element whose
singleton has zero probability can be simply thrown away from the
beginning.

It follows, than, that CPS has no null sets and, conseguentially,
typical properties are simply the holding properties.
\begin{example}
\end{example}
SOME TYPICAL PROPERTY OF THE UNBIASED UNITARY REAL SEGMENT:

The \emph{unbiased real unitary segment} is the  classical
probability space $ (  \, [  0 , 1 ) \, , , \mu_{LEBESGUE} \, )  $.

All the following predicates are clearly typical:
\begin{itemize}
  \item $ p^{1} ( x )  \; \equiv \; << \; x \, \notin \, \mathbb{Z}
  \; >> $
  \item $ p^{2} ( x )  \; \equiv \; << \; x \; \text{is transcendental}
  \; >> $
  \item $ p^{3} ( x )  \; \equiv \; << \; x \, \notin \, \mathbb{Q}
  \; >> $
\end{itemize}
\begin{example} \label{ex:some typical property of the unbiased of cbit's sequences}
\end{example}
SOME TYPICAL PROPERTY OF THE UNBIASED SPACE OF CBITS' SEQUENCES:

The \emph{unbiased real unitary segment} $ ( [ 0 ,1 ) ,
\mu_{Lebesgue} ) $ and the \emph{unbiased space of cbits'
sequences} $ \mathbf{UCS \; := \; ( \Sigma^{\infty} ,
P_{unbiased} )} $ are isomorphic as can be immediately proved
considering the \emph{dyadic expansion} of any $ x \, \in \, [ \,
0 \, , \, 1 ) $ \cite{Billingsley-95}.

Clearly such an isomorphism  maps the typical predicates $  p^{1}
\, , p^{2} \, , \, p^{3} $ of $ (  \, [  0 , 1 ) \, ,
\mu_{LEBESGUE} ) $ in typical properties $  \tilde{p}^{1} \, ,
\tilde{p}^{2} \, , \, \tilde{p}^{3} $ of $ ( \, \Sigma^{\infty}
\, ,   P_{unbiased} \, ) $.

\bigskip

Kolmogorov's original idea about the characterization of the
\textbf{intrinsic randomness} of an \textbf{individual object}
was to consider it as more random as more it is conformistic, in
the sense of conforming itself to the collectivity belonging to
all the overwhelming majorities, i.e. possesing all the typical
properties \cite{Li-Vitanyi-97}, \cite{Calude-94} .

Such an attitude results in the following:
\begin{definition}
\end{definition}
SET OF THE KOLMOGOROV-RANDOM ELEMENT OF UCS :
\begin{multline}
  KOLMOGOROV-RANDOM ( UCS ) \; \equiv \\
   \{ \, x \, \in \, M \, : p ( x ) \; holds \; \; \forall p \in {\mathcal{P}}  (  UCS
  )_{TYPICAL} \; \}
\end{multline}
But here here comes the following astonishig fact:
\begin{theorem} \label{th:not existence of Kolmogorov random sequences of cbits}
\end{theorem}
NOT EXISTENCE OF KOLMOGOROV RANDOM SEQUENCES OF CBITS
\begin{equation}
KOLMOGOROV-RANDOM ( UCS ) \; = \; \emptyset
\end{equation}
\begin{proof}
Let us introduce the following family of unary predicates over $
\Sigma^{\infty} $ depending on the parameter $ y \in
\Sigma^{\infty} $ :
\begin{equation}
  p_{y} (x) \; \equiv \; << \, x \, \neq \, y \, >>
\end{equation}
Clearly:
\begin{equation}
   p_{y} (x) \; \in \; {\mathcal{P}}  (  UCS  )_{TYPICAL} \; \;
   \forall y \in \Sigma^{\infty}
\end{equation}
and:
\begin{equation}
   p_{x} (x) \; \text{doesn't hold} \; \; \forall x \in \Sigma^{\infty}
\end{equation}
So $ p ( x ) $ is a typical property that is not satisfied by any
element of $ \Sigma^{\infty} $, immediately implying the thesis
\end{proof}

\bigskip

The theorem\ref{th:not existence of Kolmogorov random sequences of
cbits} shows that we have to relax the condition that a random
sequence of cbits possesses \textbf{all the typical properties}
requiring only that it satisfies \textbf{a proper subclass of
typical properties}.

The right subclass was proposed by P. Martin L\"{o}f  who observed
that all the Classical Laws of Randomness, i.e. all the properties
of Classical Probability Theory that are known to hold with
probability one ( such as the \textit{Law of Large Numbers}, the
\textit{Law of Iterated Logarithm} and so on ) are
\textbf{effectively-falsificable} in the sense that we can
effectively test whether they are violated ( though we cannot
effectively certify that they are satisfied).

This leads, assuming Church's Thesis \cite{Odifreddi-89} and
endowed $ \Sigma^{\infty} $ with the \textbf{product topology}
induced by the \textbf{discrete topology} of $ \Sigma $, to
introduce the following notions:

\begin{definition}
\end{definition}
$S \; \subset \; \Sigma^{\infty} $ IS ALGORITHMICALLY-OPEN:
\begin{equation}
  ( S \text{ is open } ) \; and \; ( S \, = \, X \Sigma^{\infty}
  \,
  {\mathbf{X \; recursively-enumerable}})
\end{equation}
\begin{definition}
\end{definition}
ALGORITHMIC SEQUENCE OF ALGORITHMICALLY-OPEN  SETS:

a sequence $ \{ S_{n} \}_{n \geq 1} $ of algorithmically open
sets $  S_{n} \; = \; X_{n} \Sigma^{\infty} $ : $ \exists X \;
\subset \; \Sigma^{\star} \times {\mathbb{N}} $
\textbf{recursively enumerable} with:
\begin{equation*}
   X_{n} \; = \; \{ \vec{x} \in \Sigma^{\star} \, : \, ( \vec{x} , n ) \in
   X \} \; \; \forall n \in {\mathbb{N}}_{+}
\end{equation*}
\begin{definition}
\end{definition}
$ S \; \subset \; \Sigma^{\infty} $  IS AN ALGORITHMICALLY-NULL
SET:

$ \exists  \{ G_{n} \}_{n \geq 1} $ algorithmic sequence of
algorithmically-open sets :
\begin{equation*}
  S \; \subset \; \cap_{n \geq 1} G_{n}
\end{equation*}
and:
\begin{equation*}
  alg - \lim_{n \rightarrow \infty} P_{unbiased} ( G_{n} ) \; = \; 0
\end{equation*}
i.e. there exist and increasing, unbounded, \textbf{recursive}
function $ f \, : \, {\mathbb{N}} \rightarrow {\mathbb{N}} $ so
that $ P_{unbiased} ( G_{n} ) \; < \; \frac{1}{2^{k}} $ whenever $
n \; \geq \; f(k) $

\begin{definition}
\end{definition}
LAWS OF RANDOMNESS
\begin{equation}
 \mathcal{L}_{randomness} \; \equiv \; \{ \, p ( \bar{x} )  \in  {\mathcal{P}} ( \Sigma^{\infty} ) \, :
   \{ \bar{x} \in \Sigma^{\infty} \, : \, p ( \bar{x} ) \text{ doesn't hold } \}
   \; \text{is an algorithmically null set} \}
\end{equation}
\begin{definition}\label{def:classical algorithmic randomness}
\end{definition}
MARTIN - L\"{O}F - SOLOVAY CHAITIN RANDOM SEQUENCES OF CBITS:
\begin{equation}
  RANDOM_{MLSC}(\Sigma^{\infty}) \; \equiv \; \{ \bar{x} \in \Sigma^{\infty} \: p ( \bar{x} ) \text{ holds } \; \forall p \in  {\mathcal{L}_{randomness}} \}
\end{equation}
The name in definition\ref{def:classical algorithmic randomness}
is justified by the fact the Martin - L\"{o}f characterization of
\textbf{classical algorithmic randomness} resulted to be
equivalent both to Solovay's algorithmic measure-theoretic one and
to Chaitin's definition as algorithmic incompressibility lying at
the heart of \textbf{Classical Algorithmic Information Theory}
\cite{Chaitin-87}, \cite{Calude-94}, \cite{Li-Vitanyi-97}.

To appreciate the difference between Mises-Wald-Church randomness
and Martin L\"{o}f-Solovay-Chaitin randomness let us introduce
the following:
\begin{definition}
\end{definition}
PROPERTY OF INFINITE RECURRENCE:
\begin{equation}
  p_{IR} ( \bar{x} ) := << cardinality \{ n \in {\mathbb{N}} \; :
  \;
\frac{ N ( a_{1} | \vec{x} (n))}{n} \; = \; \frac{1}{2} \} \; =
\; \aleph_{0} >>
\end{equation}
$ p_{IR} ( \cdot ) $ may be easiliy shown to be a \textit{law of
randomness}.

Anyway in 1939 J. Ville proved the following \cite{Li-Vitanyi-97}:
\begin{theorem}\label{th:Ville's theorem}
\end{theorem}
VILLE'S THEOREM

\begin{equation}
  \{ \bar{x} \in RANDOM_{MWC} \; : \; p_{IR} ( \bar{x} ) \text{ holds
  } \} \; \neq \; \emptyset
\end{equation}

Thus there exist Mises-Wald-Church random sequences that if could
occur at Casino when we are playing  as in the example\ref{ex:bet
on the last result} would give rise to the following curious
situation:
\begin{equation}
  PAYOFF(n)  \geq 0 \forall n \in {\mathbb{N}}   \; and \;  \lim_{n \rightarrow
  \infty} PAYOFF(n)  = 0
\end{equation}
i.e. , though satisfying the Law of Exluded Classical Gambling
Systems, would make us \textit{not losers} for any finite time.

Obviously, not being Martin L\"{o}f-Solovay-Chaitin random, they
won't occur with certainty.

\newpage
\section{Quantum Gambling in the framework of Quantum Statistical Decision Theory}
Quantum Decision Theory was invented by P.A. Benioff
\cite{Benioff-72} and extensively developed by C.W. Helstrom
\cite{Helstrom-76} \cite{Auletta-00}. A renewed interest in such
field has recentely grown up in the framework of Quantum Game
Theory \cite{Eisert-Wilkens-Lewenstein-99}, \cite{Boukas-00}.

As in the classical case we can always interpret a quantum
decision problem as a quantum gambling situation, with the
utility function playing the rule of the payoff.

Let us consider a gambler going to a \textbf{\textit{Quantum
Casino}} in which the croupier, at each turn n , \textbf{throws a
quantum coin}.

Such a situation may be interpreted in different ways giving rise
to different types of \textit{Quantum Casinos}.
\begin{definition} \label{def:first kind quantum casino}
\end{definition}
FIRST KIND QUANTUM CASINO:

a quantum casino specified by the following rules:
\begin{enumerate}
  \item At each turn n the croupier extracts  with unbiased
probability a \textbf{pure state} $ | \psi > (n) \; \in \;
\mathbb{H}_{2} $ , where $  \mathbb{H}_{2} $ is the \textbf{one
qubit Hilbert space}.
  \item
Before each quantum coin toss the gambler can decide, according
to a \textbf{direct gambling strategy}, among the following
possibilities:

\begin{itemize}
  \item to bet one fiche on a vector $ | \alpha > \in
  \mathbb{H}_{2} $
  \item not to bet at the turn
\end{itemize}
  \item If he decides for the first option it will happens that:
\begin{itemize}
  \item he wins a fiche if the distance among $  | \psi > (n)  $ and $
  | \alpha > $ is less or equal to fixed quantity $
  \epsilon_{Casino} $.
  \item he loses the betted fiche if the  distance among $ | \psi > (n)   $ and $
  | \alpha > $ is greater than $ \epsilon_{Casino} $
\end{itemize}
\end{enumerate}

But it is also possible to see the result of a quantum coin toss
as a \textbf{mixed state}, resulting in the following:

\begin{definition} \label{def:second kind quantum casino}
\end{definition}
SECOND KIND QUANTUM CASINO:

a quantum casino specified by the following rules:
\begin{enumerate}
  \item At each turn n the croupier extracts  with unbiased (quantum) probability a density matrix $ \rho_{n} $ on the
\textbf{one qubit alphabet} $ \mathbb{H}_{2} $.

I will denote from here and beyond with $ \mathcal{D} ( {
\mathbb{H}} ) $ the set of density matrices on the Hilbert space $
\mathbb{H} $.

  \item Before each quantum coin toss the gambler can decide, according to a \textbf{direct gambling strategy}, among the
following possibilities:

\begin{itemize}
  \item to bet one fiche on a density matrix $ \sigma \in
  \mathbb{H}_{2} $
  \item not to bet at the turn
\end{itemize}
  \item If he decides for the first option it will happens that:
\begin{itemize}
  \item he wins a fiche if the distance among $ \sigma  $ and $
  \rho_{n}$ is less or equal to fixed quantity $
  \epsilon_{Casino} $
  \item he loses the betted fiche if the  distance among $ \sigma  $ and $
  \rho_{n}$ is greater than $ \epsilon_{Casino} $
\end{itemize}
\end{enumerate}

To complete the definition of first and second kind quantum
casinos  (definition\ref{def:first kind quantum casino} and
definition\ref{def:second kind quantum casino}) we have to
clarify:
\begin{enumerate}
  \item what we mean by the distance of pure and mixed states on
an Hilbert space $ \mathbb{H} $.
  \item what we mean by a \textbf{direct gambling strategy}
\end{enumerate}

\bigskip

The more physical notions of distance between quantum states are
the following\cite{Nielsen-Chuang-00}:
\begin{enumerate}
  \item
\begin{definition} \label{def:naife quantum trace distance}
\end{definition}
QUANTUM TRACE DISTANCE ON $ \mathcal{D} ( { \mathbb{H}} ) $:
\begin{equation}
  D ( \rho_{1} , \rho_{2} ) \; := \; \frac{1}{2} \| \rho_{1} -
  \rho_{2} \|_{1} \; = \; \frac{1}{2} Tr (| \rho_{1} - \rho_{2} |
  )
\end{equation}
with:
\begin{equation}
  | a | \; := \; \sqrt{ a^{\dagger} a } \; \; a \in B ({\mathbb{H}})
\end{equation}
and where given $ n \in \{ 1 , 2 , \cdots , \infty \} $ and
denoted by $ B({\mathbb{H}}) $ the Von Neumann algebra  of the
bounded linear operators on $ \mathbb{H} $:
\begin{definition} \label{def:sequence of operatorial norms}
\end{definition}
$n^{th}$ OPERATORIAL NORM ON $ B ({\mathbb{H}}) $
\begin{equation}
  \| a \|_{n} \; := \; (Tr ( a^{\dagger}
  a)^{\frac{n}{2}}) ^{\frac{1}{n}}
\end{equation}
The $ n = \infty $ norm is called the \textbf{operator norm} and
will be considered from here and beyond as the default norm on $
B ({\mathbb{H}}) $:
\begin{equation}
  \| a \| \; := \; \| a \|_{\infty} \; \; a \in B ({\mathbb{H}})
\end{equation}

  \item

\begin{definition} \label{def:naife quantum angle distance}
\end{definition}
QUANTUM ANGLE DISTANCE ON $ \mathcal{D} ( { \mathbb{H}} ) $:

\begin{equation}
   A ( \rho_{1} , \rho_{2} ) \; := \;  \arccos F ( \rho_{1} ,
   \rho_{2} )
\end{equation}
where:
\begin{definition} \label{def:naife quantum fidelity}
\end{definition}
QUANTUM FIDELITY ON $ \mathcal{D} ( { \mathbb{H}} ) $:
\begin{equation}
   F ( \rho_{1} , \rho_{2} ) \; := \; Tr \sqrt{ \sqrt{\rho_{1}} \rho_{2} \sqrt{\rho_{1}} }
\end{equation}
\end{enumerate}
As will become clearer in the general mathematical framework of
\textbf{Quantum Probability Theory} \cite{Thirring-81},
\cite{Thirring-83}, \cite{Accardi-Frigerio-Lewis-82},
\cite{Maassen-94}, \cite{Ohya-Petz-93},\cite{Parthasarathy-92},
\cite{Meyer-95} we will  introduce later,  the
definition\ref{def:naife quantum trace distance} is the natural
quantum corrispective of the following distance on the set $
\mathcal{D} ( \Sigma ) $ of the probability distributions on a
discrete set $ \Sigma $:
\begin{definition} \label{def:naife classical trace distance}
\end{definition}
CLASSICAL TRACE DISTANCE ON $ \mathcal{D} ( \Sigma ) $:
\begin{equation}
  D ( p_{1} , p_{2} ) \; := \; \frac{1}{2} \sum_{x \in \Sigma} (  | p_{1}(x) -
  p_{2}(x) | )
\end{equation}
The intuitive meaning of the definition\ref{def:naife classical
trace distance} is clarified by the
following\cite{Nielsen-Chuang-00}:
\begin{theorem} \label{th:physical meaning of the naife classical
trace distance}
\end{theorem}
CLASSICAL TRACE DISTANCE AS DISTANCE OF THE CLASSICAL PROBABILITY
OF ANTIPODAL EVENTS:
\begin{equation}
  D ( p_{1} , p_{2} ) \; = \; \max_{e \in 2^{\Sigma}} | p_{1} (e)
  \, - \, p_{2} (e) |
\end{equation}
It is remarkable that an analogous interpretation is admissible
also in the quantum case \cite{Nielsen-Chuang-00}:
\begin{theorem} \label{th:physical meaning of the naife quantum
trace distance}
\end{theorem}
QUANTUM TRACE DISTANCE AS DISTANCE OF THE QUANTUM PROBABILITY OF
ANTIPODAL EVENTS:
\begin{equation}
  D ( \rho_{1} , \rho_{2} )  \; = \; \max_{P \in B({\mathbb{H}})_{+} }
  Tr P ( \rho_{1} - \rho_{2} )
\end{equation}
where $ B({\mathbb{H}})_{+} $ denotes the positive cone of the Von
Neumann algebra $ B({\mathbb{H}}) $ of the bounded linear
operators on $ \mathbb{H} $, i.e. the set of positive bounded
linear operators on $ \mathbb{H} $.

Theorem\ref{th:physical meaning of the naife quantum trace
distance} has an immediate interpretation in terms of quantum
measurements:
\begin{definition} \label{def:naife quantum measurement}
\end{definition}
QUANTUM MEASUREMENT ON A SYSTEM S:

a dynamical evolution of the system S, to be \textbf{necessarily}
(owed to the \textbf{\textit{endophysics - incompleteness}} of
Quantum Theory \cite{Primas-92}, \cite{Svozil-93},
\cite{Peres-95}, \cite{Rossler-98} not allowing a
\textit{description from inside} of its own means of
verification) described as an \textbf{open system}, specified by
a collection $ \{ M_{r}\}_{r \in R} $ of \textbf{trace preserving
quantum operations} on S's Hilbert space, where R is the set of
the possible measurement outcomes, such that:
\begin{equation}
  \sum_{r \in R} M_{r}^{\dagger} M_{r} \; = \; \mathbb{I}
\end{equation}
If the state of S before the measurement is $ | \psi > $ :
\begin{itemize}
  \item the probability that result $ r \in R $  occurs is given
  by:
\begin{equation}
  p(r) \; = \; < \psi | M_{r}^{\dagger} M_{r} | \psi >
\end{equation}
  \item If the result $ r \in R $ occurs, then the state of S
  after the measurement is:
\begin{equation}
  \frac{ M_{r} | \psi >  }{ \sqrt{ < \psi | M_{r}^{\dagger} M_{r} | \psi > }  }
\end{equation}
\end{itemize}

The collection of \textbf{trace preserving quantum operations} $
\{ M_{r}\}_{r \in R} $  specifying a quantum measurement induces
the following \textbf{positive operator valued measure (POVM)}  $
\{ P_{r} \}_{r \in R} $:
\begin{equation}
   P_{r} \; := \; M_{r}^{\dagger} M_{r} \; \; \forall r \in R
\end{equation}

Furhermore it may be shown that the POVM $ \{ P_{r} \}_{r \in R}
$ identifies univoquely the set of \textbf{trace preserving
quantum operations} $ \{ M_{r}\}_{r \in R} $ and can,
conseguentially, be seen as a different characterization of the
same quantum measurement.

Theorem\ref{th:physical meaning of the naife quantum trace
distance} states that $ D ( \rho_{1} , \rho_{2} ) $ is the
maximal distance of the classical probabilities of a measurement
outcome between the case in which the state before the
measurement is $ \rho_{1} $ and the case in which the state
before the measurement is $ \rho_{2} $.

Also the \textit{quantum angle distance} is the natural quantum
corrispective of a distance on the set $ \mathcal{D} ( \Sigma ) $
of the probability distributions on a discrete set $ \Sigma $,
namely:
\begin{definition} \label{def:naife classical angle distance}
\end{definition}
CLASSICAL ANGLE DISTANCE ON $ \mathcal{D} ( \Sigma ) $:
\begin{equation}
  D ( p_{1} , p_{2} ) \; := \; \arccos F ( p_{1} , p_{2} )
\end{equation}
where:
\begin{definition} \label{def:naife classical fidelity}
\end{definition}
CLASSICAL FIDELITY ON $ \mathcal{D} ( \Sigma ) $:
\begin{equation}
   F ( p_{1} , p_{2} ) \; := \;  \sum_{x \in \Sigma} \sqrt{p_{1}(x) \,  p_{2}(x)  }
\end{equation}

The name of both the  classical and the quantum angle distances
is owed to their interpretation as the angle betweeen two points
of the unit sphere: in  the classical case such interpretation is
self-evident as it appears considering the two distributions as
versors $ \begin{pmatrix}
  p_{1} ( x_{1} )   \\
  p_{1} ( x_{2} )   \\
  \vdots
\end{pmatrix} \; , \; \begin{pmatrix}
  p_{2} ( x_{1} )  \\
  p_{2} ( x_{2} )  \\
  \vdots
\end{pmatrix} $

where $  x_{1}  ,  x_{2}  ,  \cdots \; \in \; \Sigma $.

In the quantum case such an interpretation arises from the
following :
\begin{theorem}
\end{theorem}
UHLMANN'S THEOREM:
\begin{hypothesis}
\end{hypothesis}
\begin{equation*}
   \rho_{1} \, , \, \, \rho_{2} \; \in \; \mathcal{D}({\mathbb{H}})
 \end{equation*}
\begin{thesis}
\end{thesis}
\begin{equation*}
  F ( \rho_{1} \, , \, \rho_{2} ) \; = \; \max_{ | \psi_{1} > \in PUR( \rho_{1} ,
  {\mathbb{H}} ) \; , \; | \psi_{2} >  \in PUR( \rho_{2} ,
  {\mathbb{H}} )} \, | < \psi_{1} | \psi_{2} > |
\end{equation*}

where, given two generical Hilbert spaces $ \mathbb{H}_{A} $ and $
\mathbb{H}_{B} $ and a density matrix $ \rho \in {\mathcal{D}}(
{\mathbb{H}}_{A})$ :
\begin{definition}
\end{definition}
PURIFICATIONS OF $ \rho $ WITH RESPECT TO $  \mathbb{H}_{B} $:
\begin{equation}
  PUR( \rho \, , \, {\mathbb{H}_{B}}) \; := \; \{ | \psi > \in
  {\mathbb{H}_{A}} \bigotimes {\mathbb{H}_{B}} \, : \,
  Tr_{{\mathbb{H}_{B}}} | \psi > \; = \;  \rho \}
\end{equation}
So the cosin of the angle distance between two density matrices is
equal to the maximum inner product between purifications of such
density matrices.

Now here comes the rub:

which among the above two distances have we to choice in the
definition\ref{def:second kind quantum casino} of  a
\textbf{second kind Quantum Casino}?

Fortunately it can be proved that the \textbf{trace distance} and
the \textbf{angle distance} are qualitatively equivalent. Namely
\cite{Nielsen-Chuang-00}:
\begin{theorem}
\end{theorem}
QUALITATIVE EQUIVALENCE OF TRACE AND ANGLE DISTANCES ON STATES:
\begin{equation}
  1 -  F ( \rho_{1} \, , \, \rho_{2} ) \; \leq  \; D ( \rho_{1} \, , \, \rho_{2}
  ) \; \leq  \; \sqrt{1 -  F ( \rho_{1} \, , \, \rho_{2} )}
\end{equation}
and, conseguentially, it doesn't matter which of them we use in
order to define a \textbf{second kind Quantum Casino}.

For the case we are interested to in which the underlying Hilbert
space is the \textbf{one qubit Hilbert space} $ {\mathbb{H}}_{2}
$, furthermore, the adoption of the \textbf{trace distance} may
be preferred since it satisfies the following:
\begin{theorem}
\end{theorem}
QUANTUM TRACE DISTANCE IN TERMS OF THE BLOCH SPHERE:
\begin{equation}
   D (  Bloch ( \vec{r}_{1} )  , Bloch ( \vec{r}_{2} ) ) \; = \; \frac{ \| \vec{r}_{1} - \vec{r}_{2} \|
   }{2} \; \; \forall \vec{r}_{1} , \vec{r}_{2} \in Ball^{(2)}
\end{equation}
with:
\begin{definition} \label{def:Bloch sphere bijection}
\end{definition}
BLOCH SPHERE BIJECTION:

$ Bloch : Ball^{(2)} \; \rightarrow \; {\mathcal{D}} ({\mathbb{H}}
_{2} )$ :
\begin{equation}
  Bloch ( \vec{r} ) \; := \; \frac{ {\mathbb{I}} \, + \, \vec{r} \cdot \vec{\sigma}}{2}
\end{equation}
where $ Ball^{(2)} \; := \; \{ \vec{r} \in {\mathbb{R}}^{3} \, :
\, \| \vec{r} \| \leq 1 \}  $ is the unit-radius 2-ball while $
\vec{\sigma} \; := \; \begin{pmatrix}
  \sigma_{x} \\
  \sigma_{y} \\
  \sigma_{z}
\end{pmatrix} $ is the vector of the Pauli matrices.

Let us observe that the extraction with unbiased probability of
an element of $ \mathcal{D}({\mathbb{H}}_{2}) $  involved in the
definition\ref{def:second kind quantum casino} may be
reconducted, through the definition\ref{def:Bloch sphere
bijection}, to the extraction of a value of uniform-distributed
random point on the  unit radius 2-ball $  Ball^{(2)} $.

\bigskip

Let us, now, clarify what we mean by a \textbf{direct gambling
strategy}.

To make his decision at the $ n^{th} $ turn, the gambler can take
in connsideration the result of all the previous n-1 quantum coin
tosses.

He can do this in two different ways:
\begin{itemize}
  \item he can think on the \textbf{direct products} of the
  previous outcomes; we will call such a strategy a \textbf{direct gambling strategy}
  \item he can think on the \textbf{tensor products} of the
  previous outcomes;  we will call such a strategy a \textbf{tensor gambling strategy}
\end{itemize}

In a \textbf{first kind} and \textbf{second kind Quantum Casino}
the gambler has to play according to a \textbf{direct gambling
strategy}.

We will introduce, later a \textbf{third kind of Quantum Casino},
in which the gambler has to play according to a \textbf{tensor
gambling strategy}.

The \textbf{direct gambling strategies} according to which the
gambler plays in a \textbf{first kind} and \textbf{second kind
Quantum Casino} will be called, respectively, \textbf{first kind}
and \textbf{second kind quantum gambling strategies} and defined
in the following way:
\begin{definition}
\end{definition}
FIRST KIND QUANTUM GAMBLING STRATEGY:

$ S : {\mathbb{H}}_{2}^{ \star} \stackrel {\circ}{\rightarrow}
{\mathbb{H}}_{2} $

\bigskip

\begin{definition}
\end{definition}
SECOND KIND QUANTUM GAMBLING STRATEGY:

$ S : \mathcal{D}({\mathbb{H}}_{2}^{ \star}) \stackrel
{\circ}{\rightarrow} \mathcal{D}({\mathbb{H}}_{2}) $

\bigskip

Let us now consider the sets $  {\mathbb{H}}_{2}^{\infty} $ and $
\mathcal{D}({\mathbb{H}}_{2})^{\infty} $ of sequences of,
respectively, \textbf{one qubit vectors} and \textbf{one qubit
density matrices}.

Our objective is to characterize two subsets $ {\mathcal{Q}}
{\mathcal{C}}ollectives^{1} \; \subset \;
{\mathbb{H}}_{2}^{\infty}  $ and  $ {\mathcal{Q}}
{\mathcal{C}}ollectives^{2} \; \subset \;
\mathcal{D}({\mathbb{H}}_{2})^{\infty} $, that we will call,
respectively, \textbf{first kind quantum collectives} and
\textbf{second kind quantum collectives}, defined by the condition
of satisfying Von Mises's axiom\ref{ax:axiom of randomness} when
the class of the \textbf{first kind quantum admissible gambling
strategies}  $ {\mathcal{Q}}{{\mathcal{S}}trategies}_{admissible}
( {\mathcal{Q}} {\mathcal{C}}ollectives^{1} ) $ and the class of
the \textbf{second kind quantum admissible gambling strategies}  $
{\mathcal{Q}}{{\mathcal{S}}trategies}_{admissible} (
{\mathcal{Q}} {\mathcal{C}}ollectives^{2} ) $ are chosen according
to a proper algorithmic-effectiveness characterization specular
to the classical one of eq.\ref{eq:Church admissible gambling
strategies}.

We arrive, conseguentially, to the following definitions:
\begin{definition}
\end{definition}
FIRST KIND QUANTUM COLLECTIVES:

$ {\mathcal{Q}} {\mathcal{C}}ollectives^{1} \; \subset \;
{\mathbb{H}}_{2}^{\infty}  $ induced by the axiom\ref{ax:axiom of
randomness} and the assumptions that the \textbf{first kind
quantum admissible gambling strategies}    are nothing but the
\textbf{quantum algorithms} on $   {\mathbb{H}}_{2}^{\infty} $:
\begin{equation}
  {\mathcal{Q}}{{\mathcal{S}}trategies}_{admissible}
( {\mathcal{Q}} {\mathcal{C}}ollectives^{1} ) \; :=
{\mathcal{Q}}-{\mathcal{A}}lgorithms ( {\mathbb{H}}_{2}^{\infty} )
\end{equation}

\begin{definition}
\end{definition}
SECOND KIND QUANTUM COLLECTIVES:

$ {\mathcal{Q}} {\mathcal{C}}ollectives^{2} \; \subset \;
\mathcal{D}({\mathbb{H}}_{2})^{\infty} $ induced by the
axiom\ref{ax:axiom of randomness} and the assumption that the
\textbf{second kind quantum admissible gambling strategies} are
nothing but the \textbf{quantum algorithms} on $
\mathcal{D}({\mathbb{H}}_{2})^{\infty}  $:
\begin{equation}
  {\mathcal{Q}}{{\mathcal{S}}trategies}_{admissible}
( {\mathcal{Q}} {\mathcal{C}}ollectives^{2} ) \; :=
{\mathcal{Q}}-{\mathcal{A}}lgorithms (
\mathcal{D}({\mathbb{H}}_{2})^{\infty} )
\end{equation}

But we are then faced to the following dramatic question:

who is the class of\textbf{quantum algorithms} ?

The answer, which ever it is, touches the extremely subtle and
controversial debate about the relation existing between
\textbf{Church's Thesis} and \textbf{Quantum Mechanics} for which
we demand to our paper \cite{Segre-00}. Using the terminologyy
therein introduced, let us recall here briefly the key points:
\begin{itemize}
  \item \textbf{Church's Thesis} doesn't imply the equality of \textbf{Physically-classical computability} ($
  C_{\Phi}$ - computability) and \textbf{Physically-quantistical computability} ($ Q_{\Phi}$ - computability ) of \textbf{Mathematically-nonclassical objects} ($
  NC_{M}$ -objects)
  \item as far as Computability of \textbf{Mathematically-nonclassical
  objects} ($NC_{M}$ -objects) is concerned the proposal of analogous theses,
 such as the Pour - El Thesis \cite{Pour-El-Richards-89},
  \cite{Pour-El-99}, playing for special classes of \textbf{Mathematically-nonclassical
  objects} the rule played for \textbf{Mathematically-classical
  objects} ($C_{M}$ -objects)  by \textbf{Church's Thesis} should be
  considered with great attention
\end{itemize}

Let us now  pass to analyze \textbf{third kind  Quantum Casinos}.

As we already announced  in such a Quantum Casino the gambler has
to play according to a \textbf{tensor gambling strategy}.

This means that he will consider as the quantum corrispective of a
string of n cbits $ \vec{x} = x_{1} \cdots x_{n} $ not the string
of vectors $ | x_{1} > \cdots  | x_{n}
> \; \in \; {\mathbb{H}}_{2} ^{n} $ but the vector $ | x_{1}
\cdots   x_{n}
> \; \in \; {\mathbb{H}}_{2} ^{ \bigotimes n} $, where:

\begin{definition} \label{def:finite tensor power of an Hilbert space}
\end{definition}
$ n^{th}$ TENSOR POWER OF THE HILBERT SPACE $ {\mathbb{H}} $:
\begin{equation}
  {\mathbb{H}}^{ \bigotimes n } \; := \; \bigotimes_{k = 1}^{n} {\mathbb{H}}
\end{equation}
In particular:
\begin{definition} \label{def:space of the quantum strings of n qubits}
\end{definition}
SPACE OF THE QUANTUM STRINGS OF n QUBITS: $ {\mathbb{H}_{2}}^{
\bigotimes n } $

\bigskip

As to the definition of the quantum analogue of the set of
strings of an arbitrary (but finite) number of cbits $ \{ 0 , 1
\}^{\star} $  the substituion of \textbf{diretc products} by
\textbf{tensor products} automatically leads to the substitution
of the \textbf{union operator} of the definition\ref{def:
classical strings on an alphabet} by the \textbf{direct sum
operator}.

So given an Hilbert space  $ {\mathbb{H}} $, we are led to the
following definition:
\begin{definition}
\end{definition}
QUANTUM STRINGS OVER $ {\mathbb{H}} $:
\begin{equation}
   {\mathbb{H}}^{\bigotimes \star} \; := \; \bigoplus_{n \in {\mathbb{N}}} {\mathbb{H}}^{\bigotimes n}
\end{equation}
and in particular:
\begin{definition} \label{def:space of the quantum strings of qubits}
\end{definition}
SPACE OF THE QUANTUM STRINGS OF QUBITS:  $
{\mathbb{H}}_{2}^{\bigotimes \star} $

\bigskip

Clearly $ {\mathbb{H}}^{\bigotimes \star} $ is nothing but the
\textbf{Fock space} associated to $  {\mathbb{H}} $.

We can, conseguentially, look for a more autentically quantistic
definition of a \textbf{quantum gambling strategy}, in which
\textbf{entanglement} is taken into account and used by the
gambler in order to maximize his payoff.

To quantify the \textbf{entanglement properties} of quantum
strings of qubits it is useful, at this point, to introduce the
concept of \textbf{Schmidt number}.

This requires, first of all, to introduce the following basic
theorem\cite{Nielsen-Chuang-00}:
\begin{theorem} \label{th:Schimdt decomposition}
\end{theorem}
SCHMIDT DECOMPOSITION
\begin{hypothesis}
\end{hypothesis}
\begin{equation*}
  {\mathbb{H}}_{A} \: , \: {\mathbb{H}}_{B} \; \text{finite dimensional Hilbert spaces}
\end{equation*}
\begin{equation*}
  | \psi > \; \in \;  {\mathbb{H}}_{A} \bigotimes {\mathbb{H}}_{B}
\end{equation*}
\begin{thesis}
\end{thesis}
There exist  $ \{  | i_{A} > \}  $ orthonormal basis of  $
{\mathbb{H}}_{A} $   and   $ \{  | i_{B} > \} $  orthonormal
basis of $ {\mathbb{H}}_{B} $ (called the \textbf{Schmidt bases}
for, respectively,  $ {\mathbb{H}}_{A} $ and  $ {\mathbb{H}}_{B}
$) so that:
\begin{equation*}
   | \psi > \; = \; \sum_{i} \lambda_{i} | i_{A} >  | i_{B} >
\end{equation*}
where the  \textbf{Schmidt coefficients} $ \{ \: \lambda_{i} \in
{\mathbb{R}}_{+} \bigcup \{ 0 \} \: \} $ satisfy the following
coindition:
\begin{equation*}
  \sum_{i} \lambda_{i}^{2} \; = \; 1
\end{equation*}

\bigskip

Let us observe that given two finite dimensional Hilbert spaces $
{\mathbb{H}}_{A} $ and $ {\mathbb{H}}_{B} $ and a vector $ | \psi
> \; \in \;  {\mathbb{H}}_{A} \bigotimes {\mathbb{H}}_{B} $ the theorem\ref{th:Schimdt decomposition}
states the existence but not the uniqueness of the Schmidt
decomposition of $ | \psi > $

Anyway it may be proved that the following quantity is
well-defined:
\begin{definition}
\end{definition}
SCHMIDT NUMBER OF $ | \psi > $:
\begin{equation*}
  n_{Schmidt} ( | \psi > ) \; := \; cardinality \{ \lambda_{i} > 0
  \}
\end{equation*}
We will say that:
\begin{definition}
\end{definition}
$ | \psi > $ IS ENTANGLED:
\begin{equation*}
  n_{Schmidt} ( | \psi > ) \;  > \; 1
\end{equation*}
Returning to our space of qubit strings let us then introduce the
following concept:
\begin{definition}
\end{definition}
DEGREE OF ENTANGLEMENT OF THE n - QUBIT STRING $ | \psi > \in
{\mathbb{H}}_{2}^{\bigotimes n} $:
\begin{equation}
   d_{entanglement} ( | \psi > ) \;  := \max_{k = 1 , \cdots , n-1} n_{Schmidt} ( | \psi > , k ) - 1
\end{equation}
where $ n_{Schmidt} ( | \psi > , k ) $ denotes the Schmidt number
of $ | \psi > $ with respect to the splitting of
${\mathbb{H}}_{2}^{\bigotimes n} $ as
${\mathbb{H}}_{2}^{\bigotimes k} \bigotimes
{\mathbb{H}}_{2}^{\bigotimes n-k} $.

And what about quantum sequences of qubits?

They will be, clearly, the protagonists of  the limit $ n
\rightarrow \infty $ under which a Law of Excluded Quantum
Gambling Strategies for autentically quantum strategies, i.e. for
\textbf{third kind Quantum Casinos}, may be conjectured to hold.

One could think that the space of quantum sequences over $
{\mathbb{H}}_{2} $ may be easily defined in terms of the
\textbf{computational bases}.
\begin{definition}
\end{definition}
COMPUTATIONAL BASIS OF $ {\mathbb{H}}_{2} $:
\begin{equation}
    {\mathbb{E}} \; := \; \{ | 0 > , | 1 >  \} \; : < 0 | 0 > = < 1 | 1
    > = 1 \; < 0 | 1 > = 0
\end{equation}
Given any positive integer number $ n \geq 3 $:
\begin{definition}
\end{definition}
COMPUTATIONAL BASIS OF $ {\mathbb{H}}_{2}^{\bigotimes n} $:
\begin{equation}
  {\mathbb{E}}_{n} \; := \; \{ \:  | \vec{x} > \, , \,
  \vec{x} \in \{ 0 , 1 \}^{n} \: \}
\end{equation}
\begin{definition}
\end{definition}
COMPUTATIONAL BASIS OF $ {\mathbb{H}}_{2}^{\bigotimes \star} $:
\begin{equation}
  {\mathbb{E}}_{\star} \; := \; \{ | \vec{x} > \, , \,
  \vec{x} \in \{ 0 , 1 \}^{\star} \: \}
\end{equation}
The generic vector of $ {\mathbb{H}}_{2}^{\bigotimes \star} $ is
then given by a  linear combination of the form $ \sum_{ \vec{x}
\in \Sigma^{\star}} c_{\vec{x}} | \vec{x} > $.

As an example let us recall the famous
\textit{Einstein-Podolsky-Rosen (EPR)  states}:
\begin{equation*}
   \frac{ | 00 >  + | 11 >  }{\sqrt{2}}
\end{equation*}
\begin{equation*}
  \frac{ | 00 >  - | 11 >  }{\sqrt{2}}
\end{equation*}
\begin{equation*}
  \frac{ | 01 >  + | 10 >  }{\sqrt{2}}
\end{equation*}
\begin{equation*}
  \frac{ | 01 >  - | 10 >  }{\sqrt{2}}
\end{equation*}
or the equally famous \textit{Greenberger-Horne-Zeilinger (GHZ)
state}:
\begin{equation*}
   \frac{ | 1100 >  - | 0011 >  }{\sqrt{2}}
\end{equation*}

So one could think to consider the set:
\begin{equation} \label{eq:naife computational basis for quantum sequences of qubits}
   {\mathbb{E}}_{\infty} \; := \; \{ | \bar{x} > \, , \,
  \bar{x} \in \{ 0 , 1 \}^{\infty} \: \}
\end{equation}
and to define the Hilbert space of qubits' sequences by the
imposition that $ {\mathbb{E}}_{\infty} $ is a basis of it, i.e.
to introduce the following notion:
\begin{equation} \label{eq:naife space of quantum sequences of qubits}
   {\mathbb{H}}_{2}^{\bigotimes \infty} \; := \; \{ \int_{ \{0 , 1\}^{\infty}} dP_{unbiased} ( \bar{x} ) c
   (\bar{x}) \, | \bar{x} > \: , \,  c
   (\bar{x}) \in {\mathbb{C}} \: \forall \bar{x} \in
   \{ 0 , 1 \}^{\infty} \}
\end{equation}

As we saw in the example\ref{ex:some typical property of the
unbiased of cbit's sequences} the classical probability spaces $
( \, \Sigma^{\infty} \, , \, P_{unbiased} \, ) $ and $ ( \, [ 0
 ,1 \, ) \, , \, \mu_{Lebesgue} \, ) $ are isomorphic.

So  $ {\mathbb{H}}_{2}^{\bigotimes \infty} $ could seem very
similar to \textbf{the space of Dirac's kets }  $
\mathcal{H}_{ket} $ generated by \textbf{position autokets} of a
quantum nonrelativistic particle living on the unitary segment $
[ 0 ,1 ) $ :
\begin{align}
  \hat{x} & | x >  \; = \; x \,  | x > \; \; x \in [ 0 ,1 ) \\
  < x_{1} & | x_{2} > \; = \; \delta ( x_{1} -  x_{2} )
\end{align}
\begin{equation}
   \int_{ [ 0 ,1 ) } d \mu_{Lebesgue}  | x > < x | \; = \; \hat{{\mathbb{I}}}
\end{equation}
One could even be tempted to introduce by analogy a
\textbf{sequence operator} $ \hat{\bar{x}} $ on $
{\mathbb{H}}_{2}^{\bigotimes \infty} $ having $
{\mathbb{E}}_{\infty} $ as \textbf{Dirac's autokets}:
\begin{align}
  \hat{\bar{x}} & | \bar{x} >  \; = \; \bar{x} \,  | \bar{x} > \; \; \bar{x} \in \Sigma^{\infty} \\
  < \bar{x}_{1} & | \bar{x}_{2} > \; = \; \delta ( \bar{x}_{1} -  \bar{x}_{2} )
\end{align}
\begin{equation}
   \int_{ \{0 , 1\}^{\infty} } d P_{unbiased}  | \bar{x} > < \bar{x} | \; = \; \hat{{\mathbb{I}}}
\end{equation}

Now it is well known that \textbf{Dirac's original bra and ket
formalism} \cite{Dirac-58} is mathematically nonrigorous
\cite{Reed-Simon-72}, \cite{Reed-Simon-75},
\cite{Glimm-Jaffe-87}: the space of kets  $ \mathcal{H}_{ket} $
is not an Hilbert space; what constitues a well-defined Hilbert
space is the set $ L^{2} ( [0,1) , \mu_{Lebesgue} ) $ of the wave
functions $ \psi_{ | \alpha > } (x) \; := \; < x | \alpha > $.

Now, If the problem  in the introduction of $
{\mathbb{H}}_{2}^{\bigotimes \infty}$ was only this one, it would
be  justified to see it as a false problem, a mathematical
pignolery with no physical counterpart behind.

Indeed, by a restyling operation involving the  formal
sophistication of looking at $ \mathcal{H}_{ket} $ as a
\textbf{rigged Hilbert space}, Dirac's bra and ket formalism may
be recasted a completelly rigorous way.

The problem we are facing, anyway, is terribly more serious and
is not only a matter of \textbf{form } but of \textbf{substance}:

it ultimatively concerns the inadeguacy of the naife-Hilbert
space formalism of the thermodynamical limit of Quantum
Statistical Mechanics, whose solution requires the introduction of
some notion of \textbf{Quantum Probability Theory}
\footnote{Great caution must be taken in handling the locutions
\textbf{Quantum Probability Theory} and \textbf{quantum
probability space} in that they are used by different schools
with different meanings. By adhering here to Luigi Accardi -
school's terminology we invite the (eventual) lectors not to make
confusion between the definition\ref{def:quantum probability
space} and Stanley P. Gudder's definition of a \textit{quantum
probability space} as a \textit{sample space} endowed with a
\textit{probability amplitude} \cite{Gudder-88} or the
lattice-theoretic definitions such as the Enrico Beltrametti -
Gianni Cassinelli's one consisting in substituting as
\textsl{halting set} of a probability measure the classical
Booelan lattice of a $ \sigma $ - algebra of subsets of the
\textit{sample space} with a generic orthomodular lattice
\cite{Beltrametti-Cassinelli-81} or Pavel Pt\'{a}k - Sylvia
Pulmannov\'{a}'s definition of a generalized probability space as
a generic couple made up by a \textit{sum-logic} endowed with a
\textit{state} \cite{Ptak-Pulmannova-91}} \cite{Thirring-81},
\cite{Thirring-83}, \cite{Accardi-Frigerio-Lewis-82},
\cite{Maassen-94}, \cite{Ohya-Petz-93}, \cite{Parthasarathy-92},
\cite{Meyer-95}.
\begin{definition} \label{def:algebraic probability space}
\end{definition}
ALGEBRAIC PROBABILITY SPACE:   $ ( \, A \, , \, \omega \, ) $
where:
\begin{itemize}
  \item A is a Von Neumann algebra
  \item $ \omega $ is a state on A
\end{itemize}
The notion of \textbf{algebraic probability space} is a
noncommutative generalization of the notion of \textbf{classical
probability space} as is implied by the following considerations:

\begin{enumerate}
  \item a generic \textbf{classical probability space}  $ ( \; X \, , \, \mu \; ) $
  may be equivalentely seen as the \textbf{abelian algebraic probability
  space} $ ( \,  L^{\infty} ( X , \mu ) \, , \, \omega_{\mu} \,)
  $, where:
\begin{equation}
\begin{split}
  \omega_{\mu} ( A ) & \in  S ( A ) \\
  \omega_{\mu} ( a ) & \; := \; \int_{X} a ( x ) d \mu (x)
\end{split}
\end{equation}
with S(A) denoting \textbf{the set of states} over A.
  \item  given a  generic \textbf{abelian algebraic probability
  space} $ ( \, A \, , \, \omega \, ) $ there exists a \textbf{classical probability
  space} $ ( \; X \, , \, \mu \; ) $ and a
 $ \star $ - isomorphism $ {\mathcal{I}}_{GELFAND} \, : A \rightarrow  L^{\infty} ( X , \mu \,) $  called the \textbf{Gelfand
 isomorphism} under which the state $ \omega  \in S(A) $
 corresponds to the state $ \omega_{\mu} \in L^{\infty} ( X , \mu )$.
\end{enumerate}

\medskip

\begin{definition} \label{def:quantum probability space}
\end{definition}
QUANTUM PROBABILITY SPACE:  a non-abelian algebraic probability
space

\medskip

Given an \textbf{algebraic random variable} $ a \in A $ on the
\textbf{algebraic  probability space} $ ( \, A \, , \, \omega \,
) $ and a number $ n \in { \mathbb{N}}_{+} $:
\begin{definition}
\end{definition}
$ n^{th}$ MOMENT OF a :
\begin{equation}
  M_{n} (a) \; \equiv \; \omega ( a^{n} )
\end{equation}
The first moment $ M_{1} (a) $ of a \emph{noncommutative random
variable} $ a \in A $ on $ ( \, A \, , \, \omega \, ) $ is
usually called its \textbf{\emph{expectation value}} and denoted
by E(a).

An other important quantity to mention is the following:
\begin{definition}
\end{definition}
VARIANCE OF a :
\begin{equation}
  Var(a) \; \equiv \; E (a^{2}) - ( E (a) )^{2}
\end{equation}
playing a rule in the following fundamental:
\begin{theorem}
\end{theorem}
THEOREM OF INDETERMINATION:
\begin{equation}
  | E ( \frac{ [ a , b ] }{2 i} ) |  \; \leq \; \sqrt{Var(a)}
  \sqrt{Var(b)} \; \; \forall a , b \in A
\end{equation}
where $ [ a , b ] \; \equiv \; a b - b a $ is the
\emph{commutator} between a and b.

\bigskip

Given a \textbf{classical set} M , let us introduce the following
terminology:
\begin{definition}
\end{definition}
M IS OF TYPE $ I_{n} $:
\begin{equation}
  cardinality( M ) \; = \; n
\end{equation}
\begin{definition}
\end{definition}
  M IS OF TYPE $ I_{\infty} $:
\begin{equation}
  cardinality( M ) \; = \; \aleph_{0}
\end{equation}
\begin{definition}
\end{definition}
  M IS OF TYPE $ II_{1} $:
\begin{equation}
  cardinality( M ) \; = \; \aleph_{1} \; and \; M \text{ admits an unbiased probability
  measure }  P_{unbiased}
\end{equation}
\begin{definition}
\end{definition}
 M IS OF TYPE $ II_{\infty} $:
\begin{equation}
  cardinality( M ) \; = \; \aleph_{1} \; and \; M \text{ doesn't admit an unbiased probability
  measure } P_{unbiased}
\end{equation}

\begin{example}
\end{example}
THE REAL AXIS AND ITS UNITARY SEGMENT

Both the whole real axis and the its unitary segment have the
continuum power:
\begin{equation}
  cardinality({\mathbb{R}}) \; = \; cardinality( [0,1) ) \; = \; \aleph_{1}
\end{equation}
The Lebesgue measure is an unbiased probability measure for the
unitary segment while it is not a  probability measure on the
whole real axis since it is not normalizable. Hence:
\begin{align}
  Type ( [0,1) & ) \; = \; II_{1} \\
  Type ( { \mathbb{R}} & ) \; = \; II_{\infty}
\end{align}

\bigskip

An analogous situation exists in the quantum case, where the rule
of a \textbf{quantum set} \footnote{Great caution must be taken
in handling the locution \textbf{quantum set} since it is used by
various authors with completelly different meanings.Our approach
consists in considerating the words \textbf{quantum} and
\textbf{noncommutative} as synonimous and adhering to the
phylosophy underlying Noncommutative (Quantum)  Geometry
according to which one starts from the Gelfand isomorphism to
introduce noncommutative (quantum) spaces and then define on them
the whole hierarchy of more and more refined structures:
measure-theoretic (the only one playing a rule in this paper),
topological, differential-geometric, and
(pseudo)riemannian-geometric.Such an acception of the locution
\textbf{quantum set} is completelly different both from the
exoteric  \textbf{quantum sets} of Takeuty
\cite{Dalla-Chiara-Giuntini-01}   and from the even more exoteric
\textbf{quantum sets} of Finkelstein
\cite{Finkelstein-97}.Finally it must be remarked that the same
expression \textbf{quantum set} may be someway misleading because
someone  could be tempted to look at it erroenously as a departure
from Classical Set Theory for a new set theory: this is absolutely
not the case since a Von Neummann algebra is, in particular,
obviously a classical set!!!} is  played by a non-abelian Von
Neumann algebra A or, better, the building blocks it is made of,
i.e. the \textbf{factors} contributing to its \textbf{factor
decomposition}.

Introduced the following terminology:
\begin{definition}
\end{definition}
QUANTUM UNBIASED PROBABILITY MEASURE ON A:  a finite, faithful,
normal trace on A

\medskip

and assumed for simplicity that A is itself a factor we have a
classification very similar to the classical one:
\begin{definition}
\end{definition}
A IS OF TYPE $ I_{n} $:
\begin{equation}
  cardinality(range(d)) \; = \; n
\end{equation}
\begin{definition}
\end{definition}
  A IS OF TYPE $ I_{\infty} $:
\begin{equation}
  cardinality(range(d)) \; = \; \aleph_{0}
\end{equation}
\begin{definition}
\end{definition}
  A IS OF TYPE $ II_{1} $:
\begin{equation}
  cardinality(range(d)) \; = \; \aleph_{1} \; and \; A \text{ admits an unbiased quantum probability
  measure } P_{unbiased}
\end{equation}
\begin{definition}
\end{definition}
 A IS OF TYPE $ II_{\infty} $:
\begin{equation}
  cardinality(( range(d))) )\; = \; \aleph_{1} \; and \; A \text{ doesn't admit an unbiased probability
  measure } P_{unbiased}
\end{equation}

where  d  is an arbitary \textbf{dimension function} on the
\textbf{complete}, \textbf{orthomodular} lattice
 Pr(A) of the projections of A \cite{Beltrametti-Cassinelli-81}, \cite{Ptak-Pulmannova-91}, \cite{Svozil-98}, \cite{Redei-98}, \cite{Dalla-Chiara-Giuntini-97}, \cite{Dalla-Chiara-Giuntini-01}.

It must be said, for completeness, that in the quantum case there
exist a suppletive one-parameter family of cases  $ Type(A) \; =
\; III_{\lambda} \, \lambda \in [0,1] $ having no corrispective
in the classical case that (fortunately) won't have no rule in
this paper.

\bigskip

These considerations justify the introduction of the following
terminology:
\begin{definition} \label{def:one qubit algebraic alphabet}
\end{definition}
ONE QUBIT ALGEBRAIC ALPHABET:
\begin{equation}
  \Sigma_{alg} \; := \; M_{2} ({\mathbb{C}})
\end{equation}
\begin{definition} \label{def:algebraic space of quantum strings of n qubits}
\end{definition}
ALGEBRAIC SPACE OF QUANTUM STRING OF n QUBITS:
\begin{equation}
   \Sigma_{alg}^{\bigotimes n } \; := \; M_{2} ({\mathbb{C}})^{\bigotimes n}
\end{equation}
\begin{definition} \label{def:algebraic space of quantum strings of qubits}
\end{definition}
ALGEBRAIC SPACE OF QUANTUM STRINGS OF QUBITS:
\begin{equation}
   \Sigma_{alg}^{\bigotimes \star } \; := \;
   \bigoplus_{{\mathbb{N}}} \Sigma_{alg}^{\bigotimes n }
\end{equation}

\medskip

Obviously $ \Sigma_{alg} $ is a $ I_{2} $-factor; furthermore $
\Sigma_{alg}^{\bigotimes n }$, being $ \star $ - isomorphic to $
M_{n} ({\mathbb{C}}) $, is a $ I_{n} $-factor. Hence it is also $
\star $ - isomorphic to $ B ( {\mathbb{H}}_{n} ) $ (with $
{\mathbb{H}}_{n} $  denoting an n - dimensional Hilbert space)
and admits the unbiased quantum probability measure $ \tau_{n} (
\cdot ) \, := \, \frac{1}{n} Tr ( \cdot ) $.

Furthermore  every state $ \omega \in S (
\Sigma_{alg}^{\bigotimes n } )$ on it is \textbf{normal} and hence
there exists a density matrix $ \rho_{\omega} \in {\mathcal{D}}
({\mathbb{H}}_{2}^{\bigotimes n}) $ so that:
\begin{equation} \label{eq: density matrix of a normal state}
  \omega ( a ) \; = \; Tr( \rho_{\omega} a ) \; \; \forall a \in \Sigma_{alg}^{\bigotimes n }
\end{equation}

Eq.\ref{eq: density matrix of a normal state} shows that the
algebraic characterization of quantum strings is absolutely
equivalent to the usual Hilbert space one  based on the
definitions definition\ref{def:space of the quantum strings of n
qubits} and definition\ref{def:space of the quantum strings of
qubits}.

The whole operator algebraic machinery would then seem (as is,
indeed, often considered by not enough accultured physicists) an
arbitrary mathematical sophistication to recast in a hieratic
mathematical language simple physical statements.

As far as our issues is concerned, if for $ n \rightarrow \infty
\; \; \Sigma_{alg}^{\bigotimes n } $ tended to an $ I_{\infty}
$-factor $ \Sigma_{alg}^{\bigotimes \infty} $ this would be
indeed true since, in this case, there would exist an
infinite-dimensional Hilbert space $ \mathbb{H} $ such that:
\begin{enumerate}
  \item $  \Sigma_{alg}^{\bigotimes \infty} $ would be  $
\star $ - isomorphic to $ B ( {\mathbb{H}}) $
  \item   every state on $ \Sigma_{alg}^{\bigotimes
\infty} \;  \omega \in S ( \Sigma_{alg}^{\bigotimes \infty }) $
would be  \textbf{normal} and hence there would exist a density
matrix $ \rho_{\omega} \in {\mathcal{D}}
({\mathbb{H}}_{2}^{\bigotimes \infty}) $ so that:
\begin{equation}
  \omega ( a ) \; = \; Tr( \rho_{\omega} a ) \; \; \forall a \in \Sigma_{alg}^{\bigotimes \infty }
\end{equation}
\end{enumerate}
admitting to recast again the analysis in the usual Hilbert space
formulation (at the price of some quantum-logical subtility owed
to the fact that the lattice $ Pr( \Sigma_{alg}^{\bigotimes
\infty } ) $ in this case wouldn't be \textbf{modular})

\textbf{But the factor $ \Sigma_{alg}^{\bigotimes \infty} $ to
which $ \Sigma_{alg}^{\bigotimes n } $ tends for $ n \rightarrow
\infty $  is of type $ II_{1} $.}

This can be shown in the following way:

the restriction of the unbiased quantum probability measure $
\tau_{n} $  to $  Pr ( \Sigma_{alg}^{\bigotimes n }) $ is a
dimension function $ d_{n} $ so that:
\begin{equation}
  Range( d_{n} ) \; = \; \{ \frac{k}{2^{n}} \, : \, k = 0
  , 1 , 2 , \cdots , 2^{n} \}
\end{equation}
Since:
\begin{equation}
  \lim_{ n \rightarrow \infty } cardinality ( \{ \frac{k}{2^{n}} \, : \, k = 0
  , 1 , 2 , \cdots , 2^{n} \} ) \;
  = \; \aleph_{1}
\end{equation}
it follows that the infinite tensor product of $ M_{2} (
{\mathbb{C}}) $ can't be of type $ I_{\infty} $ and,
conseguentially:
\begin{enumerate}
  \item it is not $ \star $-isomorphic to a $ B ( {\mathbb{H}}) $
  \item  a state on it is not, in general, normal and, hence,
  can't be represented by a density matrix
\end{enumerate}
Making more rigorous the previous informal arguments let us
introduce the following:
\begin{definition} \label{def:algebraic space of quantum sequences of qubits}
\end{definition}
ALGEBRAIC SPACE OF QUANTUM SEQUENCES OF QUBITS:
\begin{equation}
   \Sigma_{alg}^{\bigotimes \infty  } \; := \; completion_{\| \, \cdot
   \|} \bigotimes_{{\mathbb{N}}} M_{2} ( {\mathbb{C}} )
\end{equation}
It can be proved that $  \Sigma_{alg}^{\bigotimes \infty  } $
admits an unbiased quntum probability measure and is then of type
$ II_{1} $(implying that the lattice $  Pr(
\Sigma_{alg}^{\bigotimes \infty  } )$ is \textbf{modular}).

\bigskip

Let us finally introduce the following notions:
\begin{definition} \label{def:algebraic quantum coin}
\end{definition}
 ALGEBRAIC QUANTUM COIN: a \textbf{quantum random variable} on the \textbf{quantum probability
 space} $ ( \, M_{2} (  {\mathbb{C}} )  \, , \, \tau_{2} ) $

 \medskip

\begin{definition} \label{def:third kind quantum casino}
\end{definition}
THIRD KIND QUANTUM CASINO:

a quantum casino specified by the following rules:
\begin{enumerate}
  \item At each turn n the croupier throws an \textbf{algebraic quantum coin} $ A_{n} $ obtaining a value $
  a_{n} \, \in \, \Sigma_{alg} $
  \item
Before each algebraic quantum coin toss the gambler can decide,
by adopting a \textbf{quantum gambling strategy}, among the
following possibilities:

\begin{itemize}
  \item to bet one fiche on an a letter  $ b  \,\in \,
  \Sigma_{alg} $
  \item not to bet at the turn
\end{itemize}
  \item If he decides for the first option it will happens that:
\begin{itemize}
  \item he wins a fiche if the distance among $  a_{n}  $ and b  $ d( a_{n} , b ) \; := \; \| a_{n} - b \| $ is less or equal to a fixed quantity $
  \epsilon_{Casino} $.
  \item he loses the betted fiche if the  distance among $  a_{n}  $  and
  a $ d( a_{n} , b ) \; := \; \| a_{n} - b \| $  is greater than $ \epsilon_{Casino} $
\end{itemize}
\end{enumerate}

\medskip

where the adoption of a \textbf{tensor gambling strategy} is
formalized in terms of the following notion:
\begin{definition} \label{def:third kind quantum gamling straregy}
\end{definition}
THIRD KIND QUANTUM GAMBLING STRATEGY:

$ S : \Sigma_{alg}^{\bigotimes \star }  \stackrel
{\circ}{\rightarrow} M_{2} ({\mathbb{C}}) $

\medskip

The concrete way in which the gambler applies, in every kind of
Quantum Casino, the chosen strategy S is always the same:
\begin{itemize}
  \item if S doesn't halt on the \textbf{previous game history} he doesn't bet
  at the next turn
  \item if S halts on on the past game history he bets  S(previous game
  history)

\end{itemize}

Lets us denote by  $ \bar{a}  \in \Sigma_{alg}^{\bigotimes
\infty} $ the occured quantum sequence of qubits and with $
\vec{a}(n) \; := a_{1} \, \bigotimes  \, \cdots \, \bigotimes \,
a_{n} \in \Sigma_{alg}^{\bigotimes n} $ its \textbf{quantum
prefix of length n}, i.e. the quantum string of the results of
the first n quantum coin tosses.

Quantum Casinos could seem , at this point, an abstruse
mathematical concept; they are, anyway, as concrete as classical
casinos and may be concretelly simulated by the following
Mathematica code:
\newpage

\includegraphics[scale=0.85]{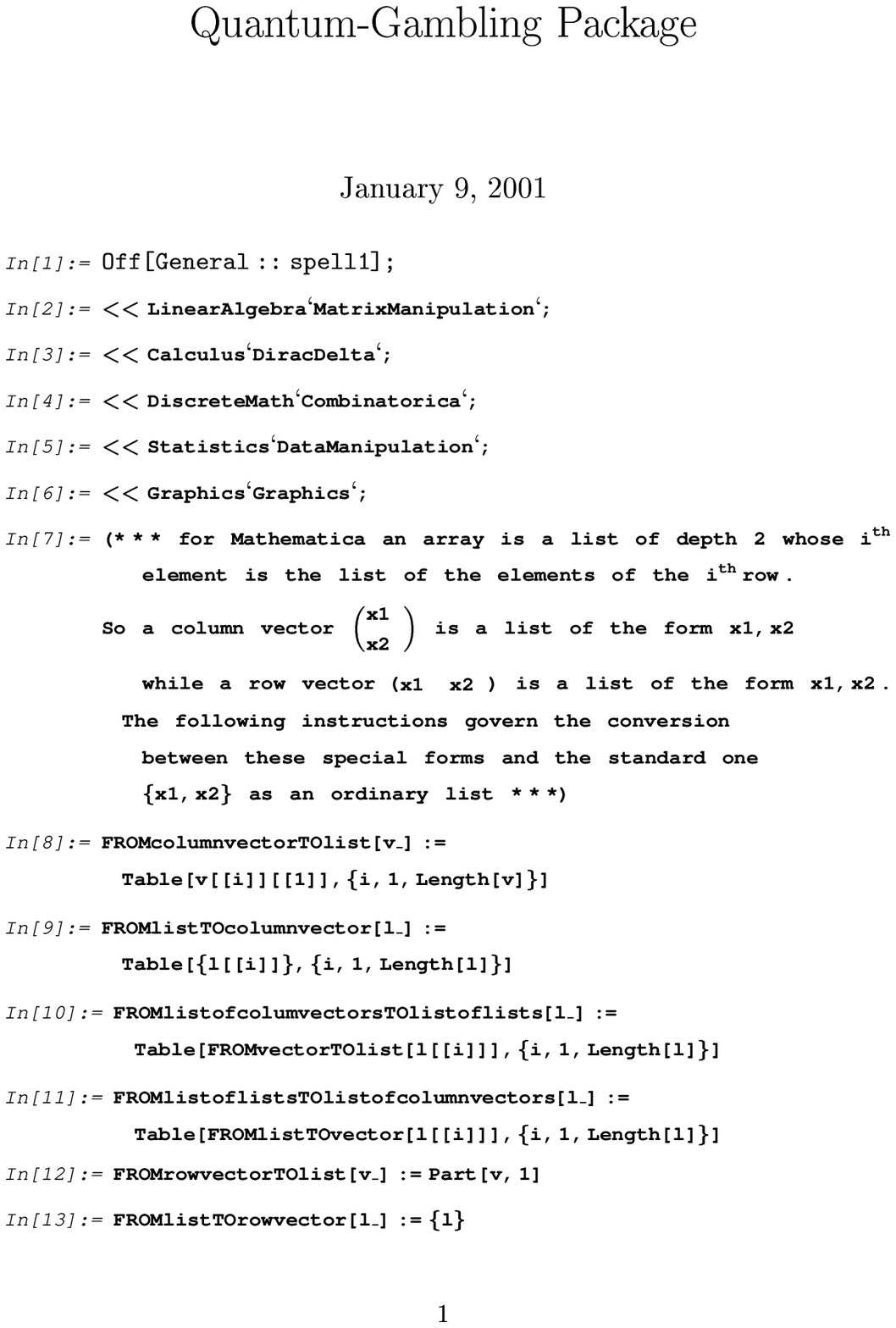}
\includegraphics[scale=0.85]{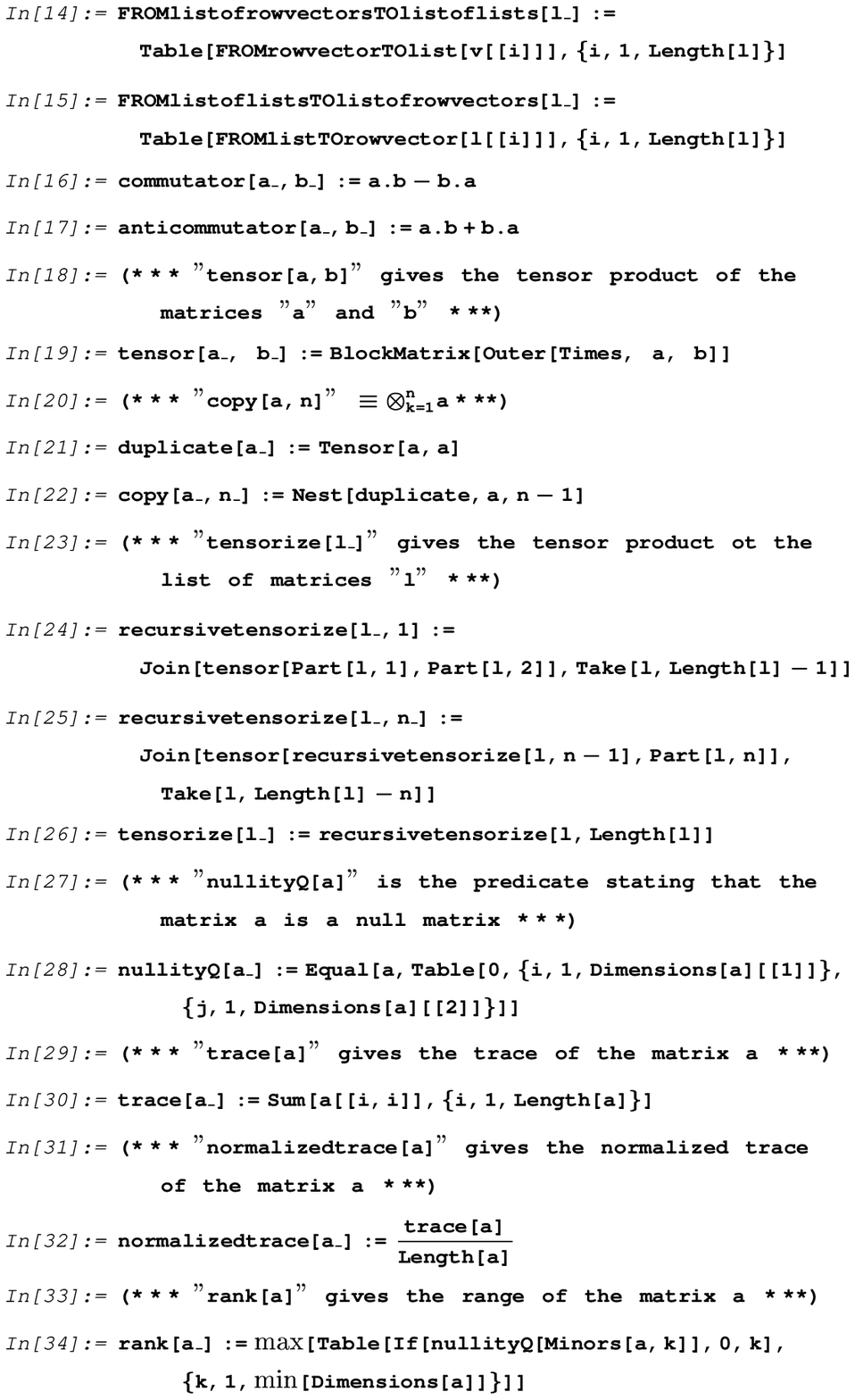}
\includegraphics[scale=0.85]{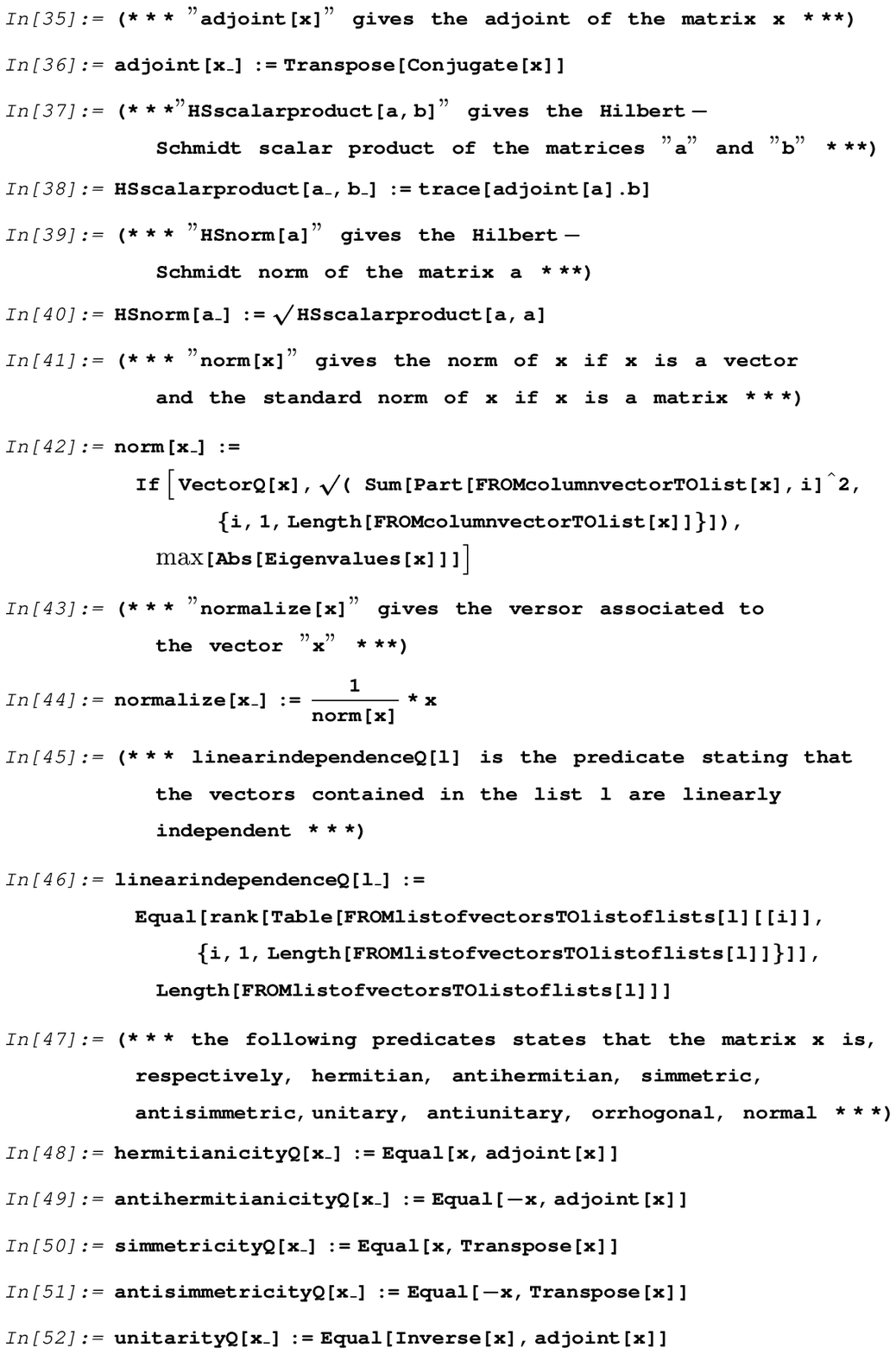}
\includegraphics[scale=0.85]{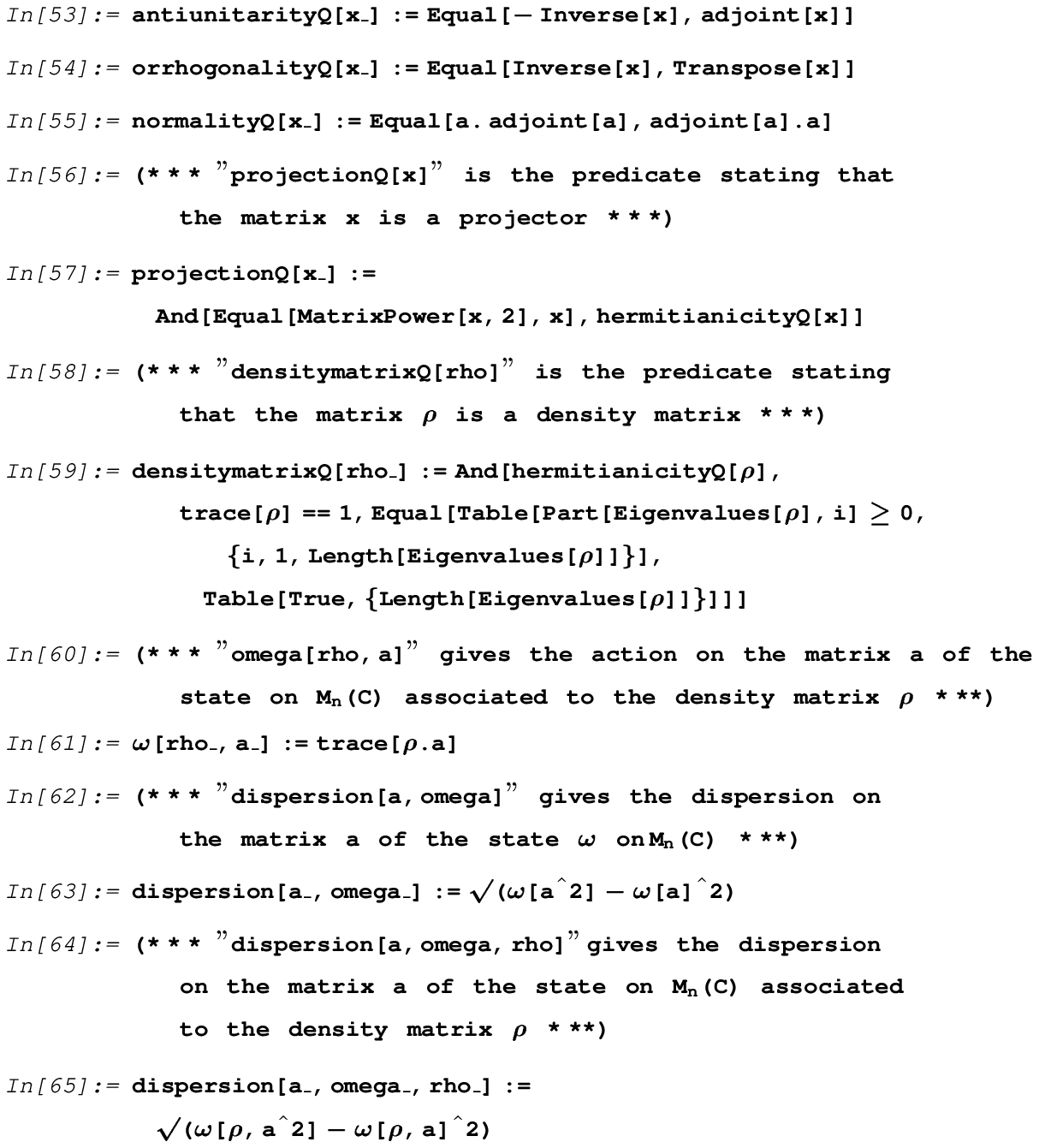}
\includegraphics[scale=0.85]{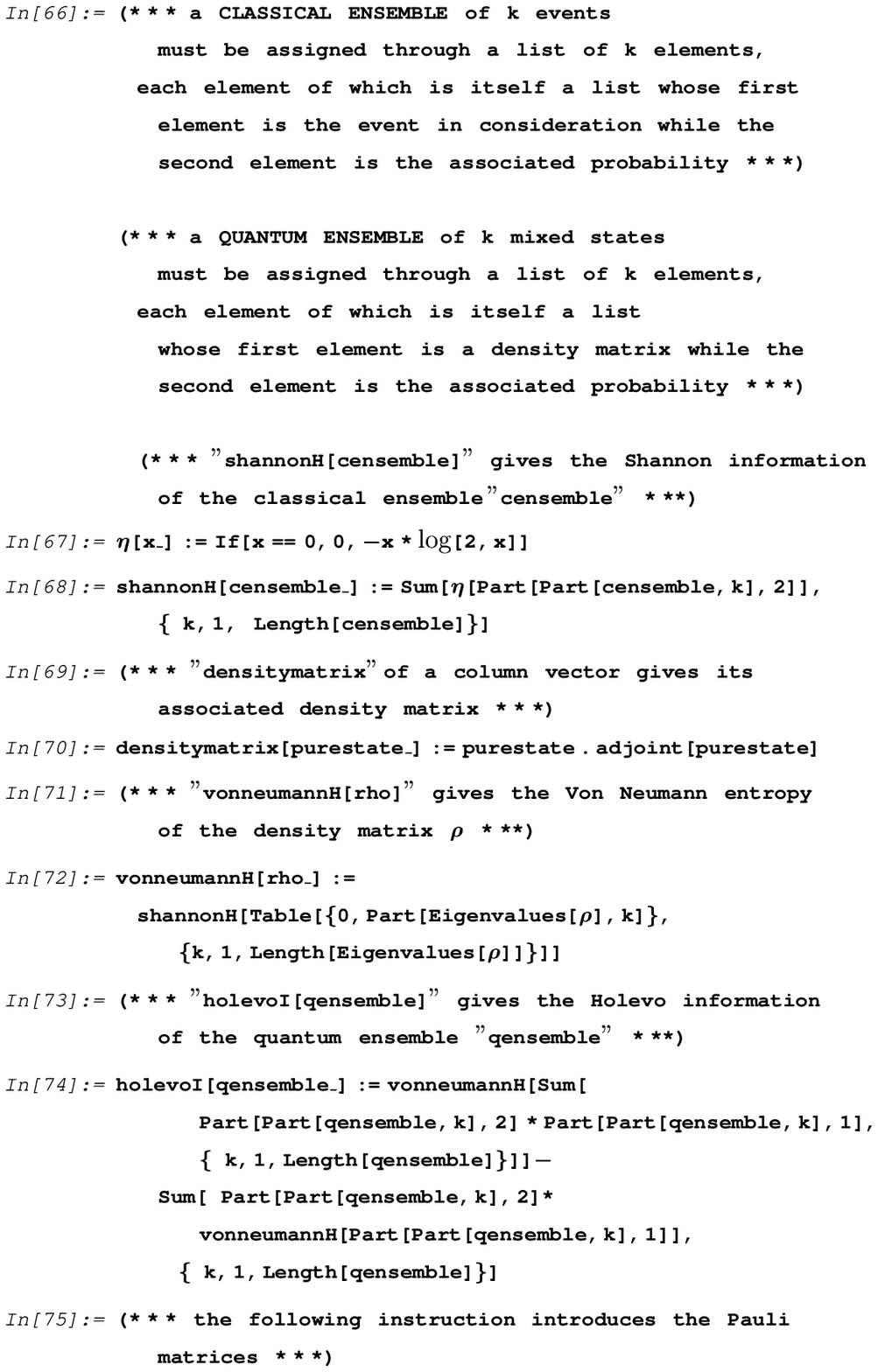}
\includegraphics[scale=0.85]{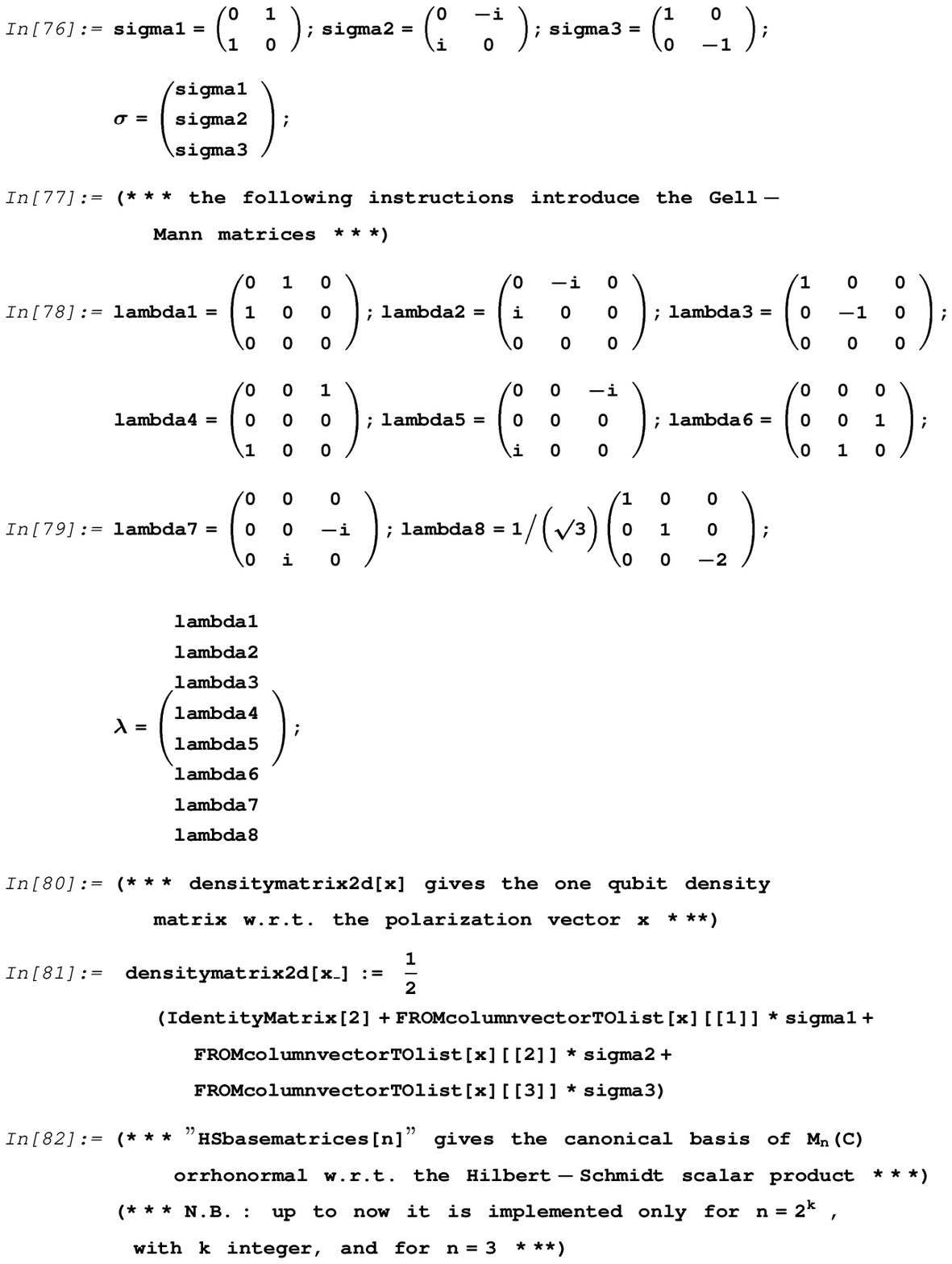}
\includegraphics[scale=0.85]{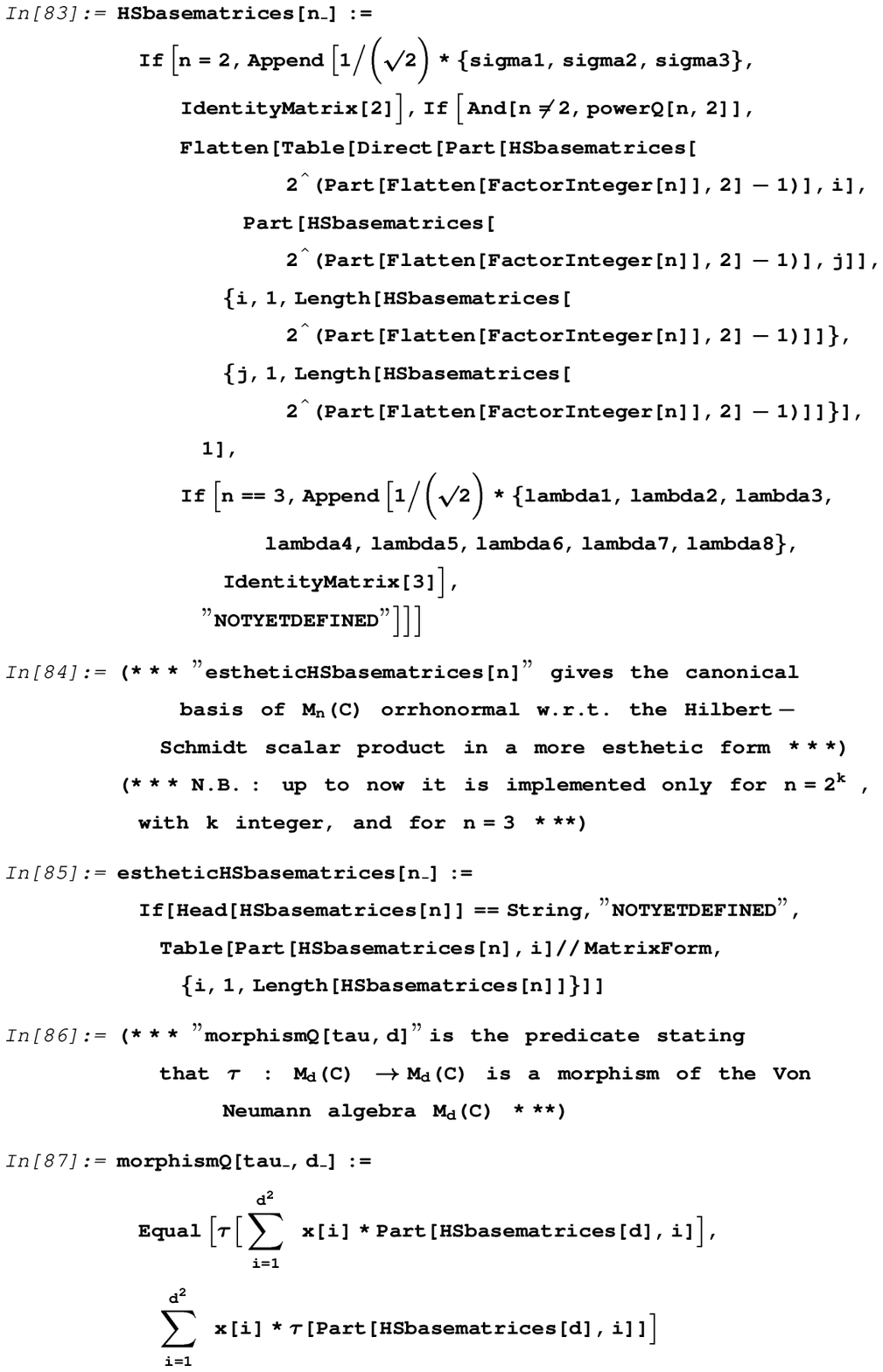}
\includegraphics[scale=0.85]{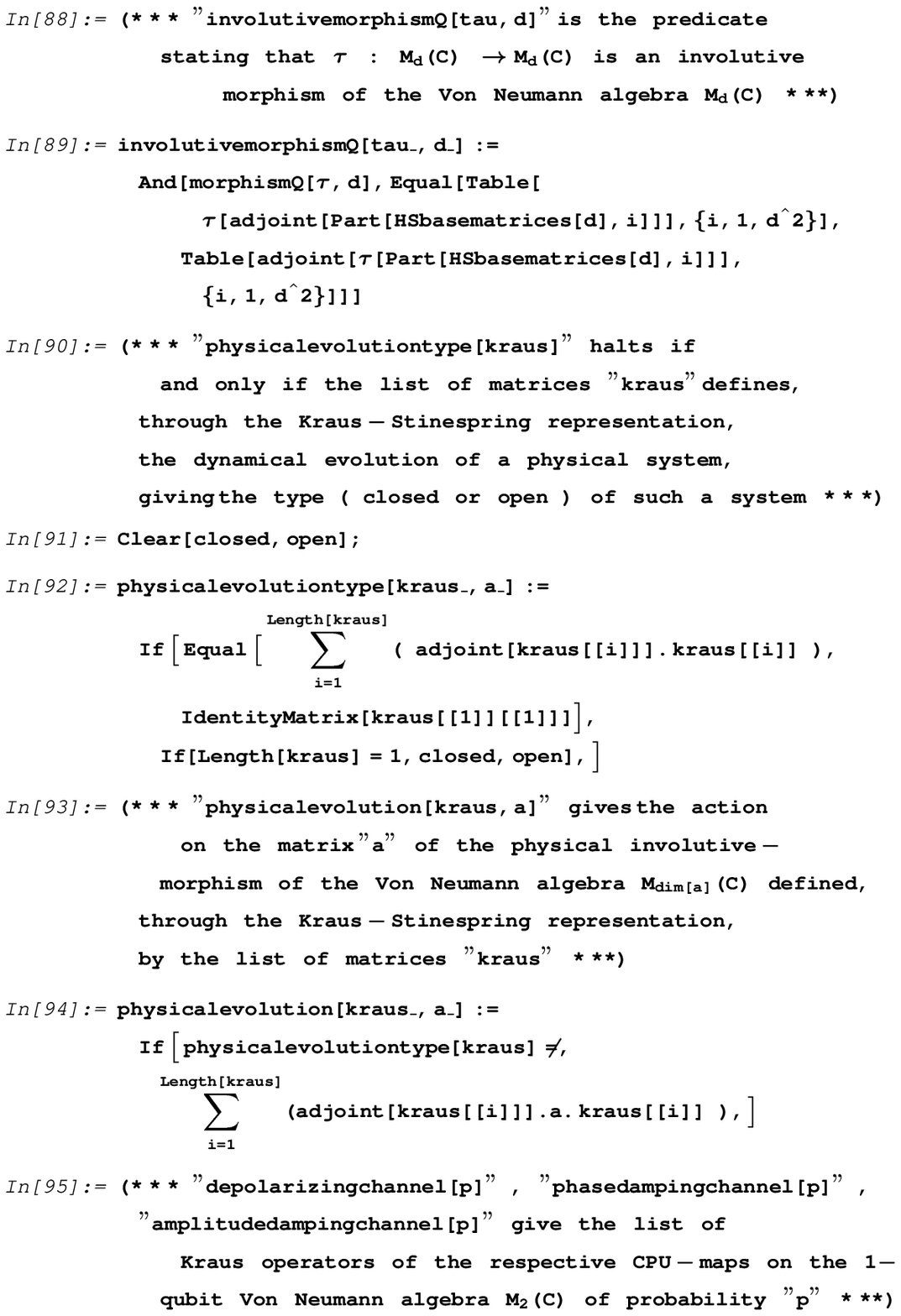}
\includegraphics[scale=0.85]{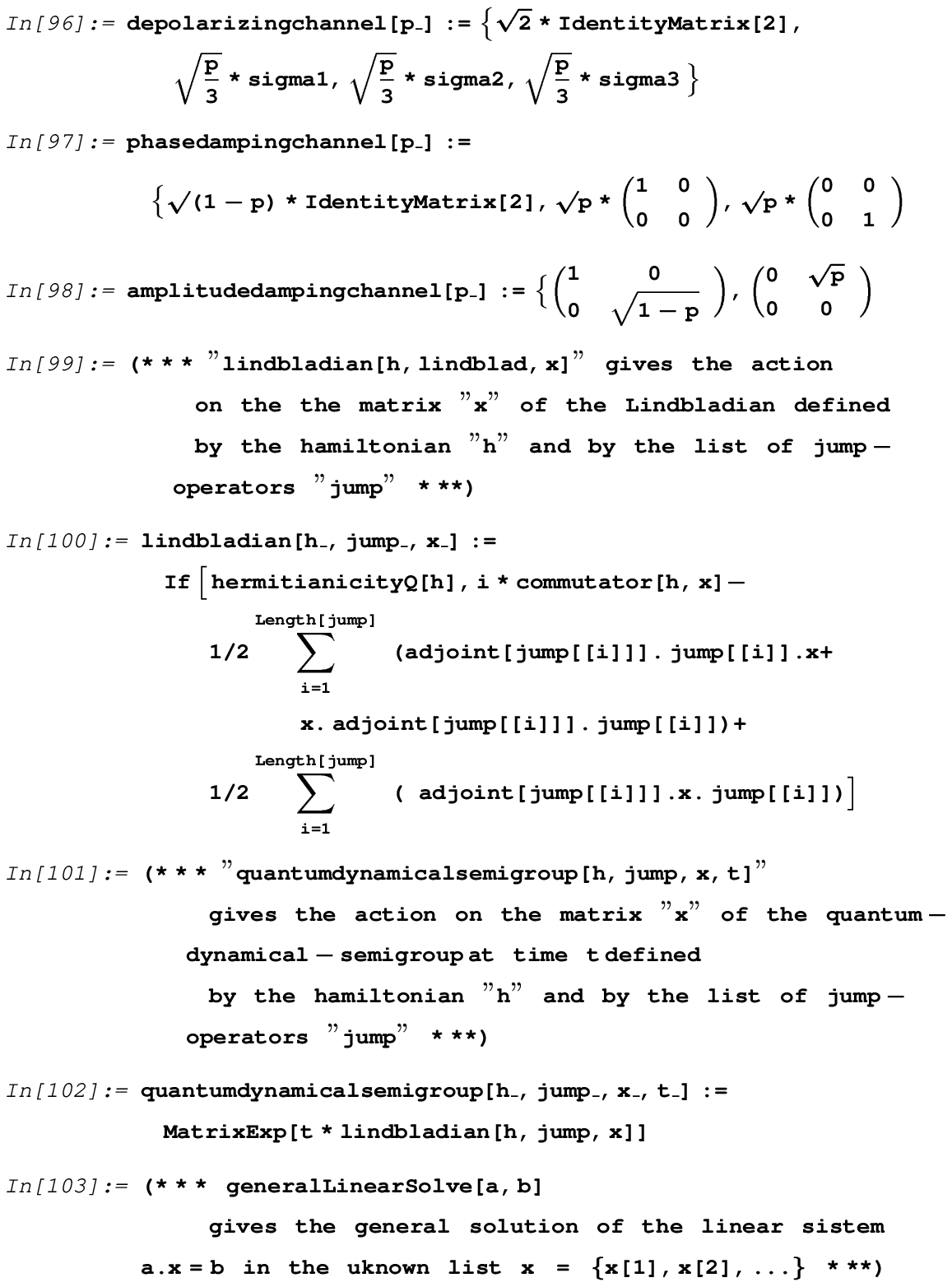}
\includegraphics[scale=0.85]{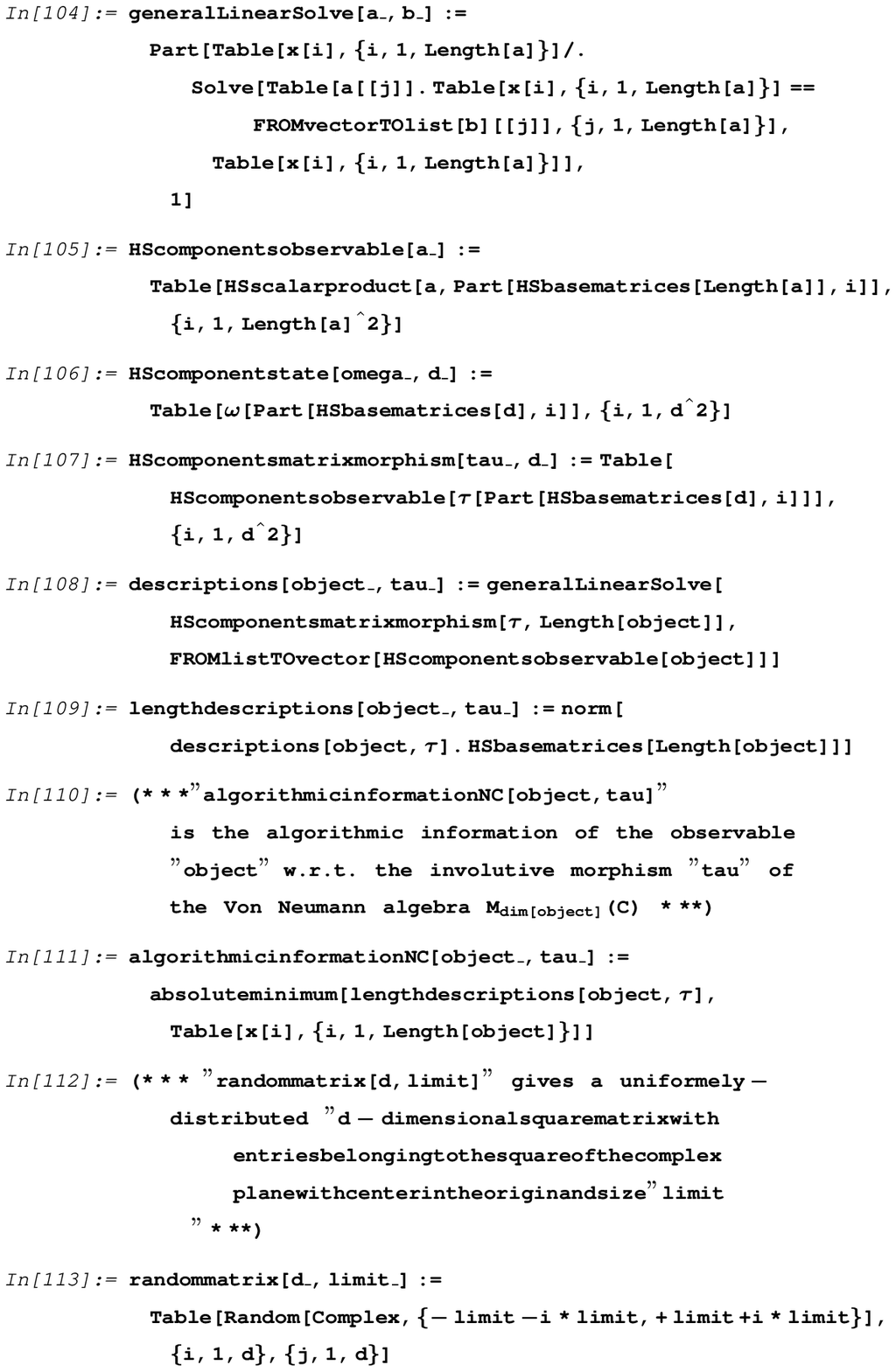}
\includegraphics[scale=0.85]{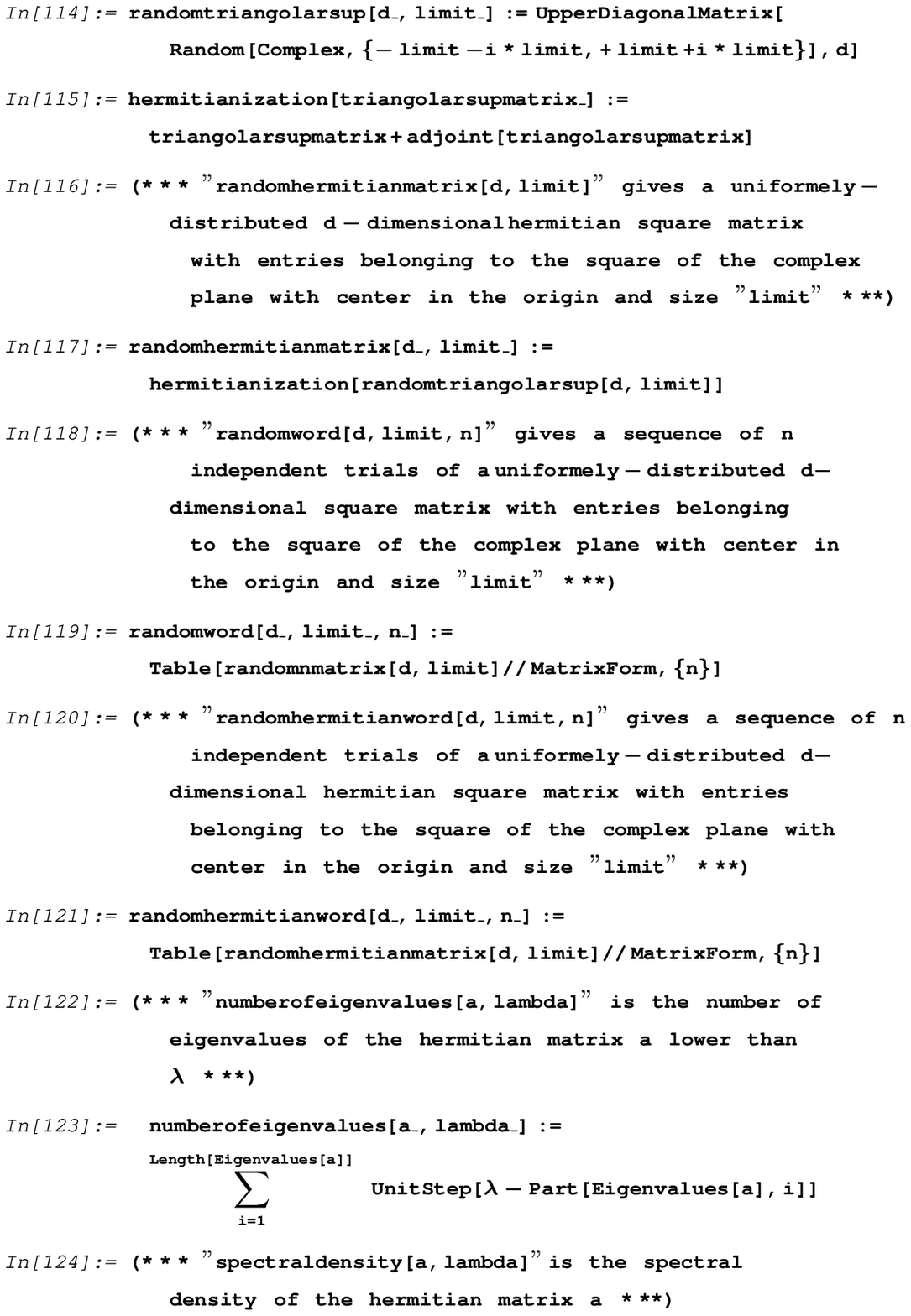}
\includegraphics[scale=0.85]{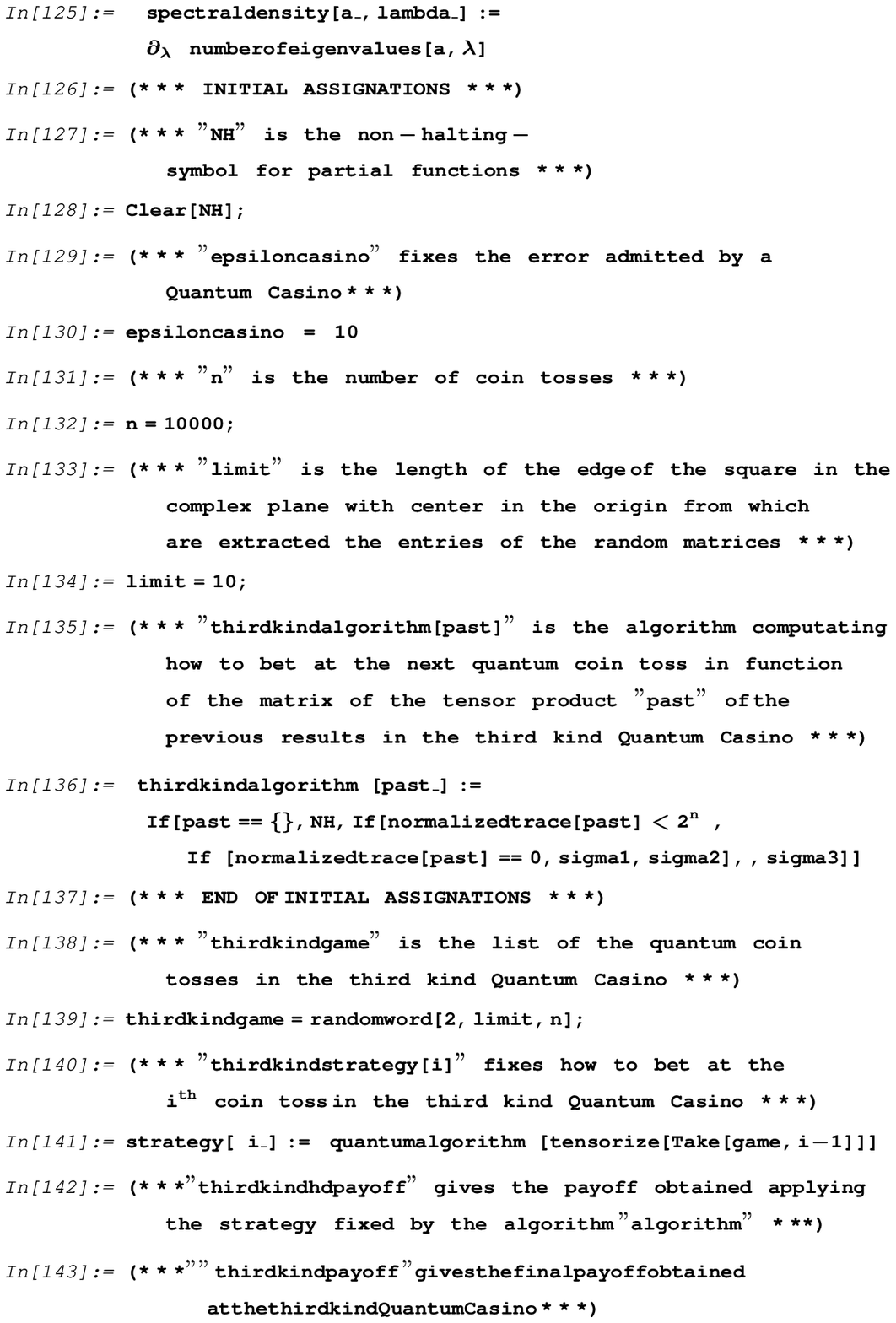}
\includegraphics[scale=0.85]{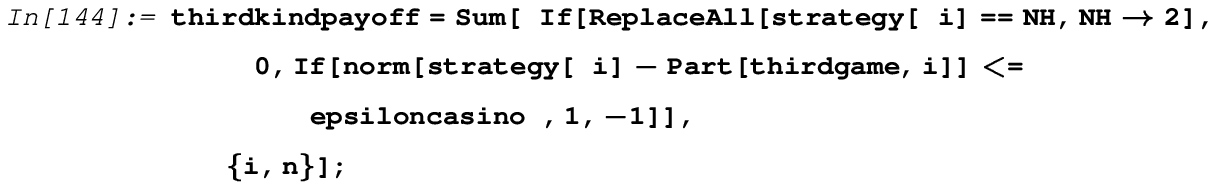}
\newpage
where the initial assignations may be arbitarily variated from
their default: a number of $ 10000 $ quantum coin tosses, an error
of $ \epsilon_{Casino} \; = \; 10 $, an edge's length of the
origin-centered square of the complex plane to which belong random
matrices' entries of 10 and the adoption of the gambling strategy
discussed in the following example:

\begin{example}
\end{example}
BETTING ON PAULI MATRICES CHOOSING ACCORDING TO THE HEIGHT OF THE
UNBIASED QUANTUM PROBABILITY MEASURE

Let us consider the following \textbf{third kind quantum gambling
strategy}:
\begin{equation}
  S ( \vec{a}(n) ) \; := \;
  \begin{cases}
    \uparrow   & \text{if $ \vec{a}(n)  = \lambda $ }, \\
    \sigma_{x} & \text{if $ P_{unbiased} ( \vec{a}(n)^{\dagger} \vec{a}(n) ) \; = \; 0 $}, \\
    \sigma_{y} & \text{if $ P_{unbiased} ( \vec{a}(n)^{\dagger} \vec{a}(n) ) < 2^{n} $}, \\
    \sigma_{z} & \text{otherwise}.
  \end{cases}
\end{equation}
where $ \lambda $ denotes the empty quantum string.

Let us imagine that the results of the first three quantum coin
tosses are:
\begin{equation*}
  a(1) \; = \; \begin{pmatrix}
    5.21295-0.543424 I & -5.83373-1.51207 I \\
    -5.72507+5.64286 I & 0.264194-5.36408 I \
  \end{pmatrix}
\end{equation*}
\begin{equation*}
   a(2) \; = \; \begin{pmatrix}
     -2.21604-8.29818 I & 2.29687-9.22925 I \\
     -7.10612+4.25443 I & -8.19842+6.03258 I \
   \end{pmatrix}
\end{equation*}
\begin{equation*}
  a(3) \; = \; \begin{pmatrix}
    9.80519-7.0523 I & -7.72367-6.40421 I \\
    -0.227234+7.87254 I & 6.36604+6.81784 I \
  \end{pmatrix}
\end{equation*}
so that:
\begin{equation*}
  \vec{a}(1) \; = \; \begin{pmatrix}
    5.21295-0.543424 I & -5.83373-1.51207 I \\
    -5.72507+5.64286 I & 0.264194-5.36408 I \
  \end{pmatrix}
\end{equation*}
\begin{multline*}
   \vec{a}(2) \; = \; a(1) \, \bigotimes \, a(2) \; = \\ \begin{pmatrix}
     -16.0615-42.0538 I & 6.95806-49.3598 I & 0.380344+51.7601 I & -27.3546+50.3679 I \\
     -34.7319+26.0397 I & -39.4597+35.9027 I & 47.8881-14.0742 I & 56.949-22.7959 I \\
     59.5124+35.0029 I & 38.9296+65.799 I & -45.0976+9.69467 I & -48.8996-14.7589 I \\
     16.6759-64.4557 I & 12.8955-80.7994 I & 20.9437+39.2418 I & 30.1933+45.5707 I \
   \end{pmatrix}
\end{multline*}
\begin{multline*}
  \vec{a}(3) \; = \; a(1) \, \bigotimes \, a(2) \; \bigotimes \,
  a(3) \; = \\
  \begin{scriptsize}
    \begin{pmatrix}
    -454-299 I & -145+428 I & -280-533 I & -370+337 I & 369+505 I & 329-402 I & 87+687 I & 534-214 I \\
    335-117 I & 184-377 I & 387+66 I & 381-267 I & -408-8.8 I & -350+332 I & -390-227 I & -518+134 I \\
    -157+500 I & 435+21 I & -134+630 I & 535-25 I & 370-476 I & -460-198 I & 398-625 I & -586-189 I \\
    -197-279 I & -399-71 I & -274-319 I & -496-41 I & 100+380 I & 401+237 I & 167+454 I & 518+243 I \\
    830-76 I & -235-651 I & 846+371 I & 121-758 I & -374+413 I & 410+214 I & -584+200 I & 283+427 I \\
    -289+461 I & 140+629 I & -527+292 I & -201+684 I & -66-357 I & -353-246 I & 127-382 I & -211-427 I \\
    -291-750 I & -542+391 I & -443-883 I & -617+541 I & 482+237 I & 90-437 I & 617+234 I & 59-545 I \\
    504+146 I & 546-297 I & 633+120 I & 633-426 I & -314+156 I & -134+393 I & -366+227 I & -118+496 I \
    \end{pmatrix}
  \end{scriptsize}
\end{multline*}
where we have passed from four to zero decimal ciphres to save
space.

Gambler's evening to a third kind quantum casino may be told in
the following way:
\begin{itemize}
  \item at the beginning he has $ PAYOFF( 0 ) \; = \; 0 $ ; since
  at the first turn he doesn't bet we have obviously that $  PAYOFF( 1 ) \; = \; 0 $
  \item since:
\begin{multline*}
  P_{un} ( \begin{pmatrix}
    5.21295-0.543424 I & -5.83373-1.51207 I \\
    -5.72507+5.64286 I & 0.264194-5.36408 I \
  \end{pmatrix} ^{\dagger} \, \begin{pmatrix}
    5.21295-0.543424 I & -5.83373-1.51207 I \\
    -5.72507+5.64286 I & 0.264194-5.36408 I \
  \end{pmatrix} ) \\
   = \; P_{un} ( \begin{pmatrix}
    5.21295+0.543424 I & -5.72507-5.64286 I \\
    -5.83373+1.51207 I & 0.264194+5.36408 I \
  \end{pmatrix} \, \begin{pmatrix}
    5.21295-0.543424 I & -5.83373-1.51207 I \\
    -5.72507+5.64286 I & 0.264194-5.36408 I \
  \end{pmatrix} )\; = \\
   P_{unbiased} ( \begin{pmatrix}
    92.0884 & -61.3705+18.1664 I \\
    -61.3705-18.1664 I & 65.1619 \
  \end{pmatrix} ) \; = \; 157.25 \; > \; 2
\end{multline*}
he bets on $ \sigma_{z} $.

 \item since:
\begin{equation*}
  \| a(2) \, - \, \sigma_{z} \| \; = \; \| \begin{pmatrix}
    -3.21604-8.29818 I & 2.29687-9.22925 I \\
    -7.10612+4.25443 I & -7.19842+6.03258 I \
  \end{pmatrix}  \|  \; = \; 11.5984 \; > \; 10
\end{equation*}
he loses his fiche. Consequentially $ PAYOFF(1) \; = \; -1 $
  \item since:
\begin{multline*}
   P_{un} ( \begin{pmatrix}
     -16.0615-42.0538 I & 6.95806-49.3598 I & 0.380344+51.7601 I & -27.3546+50.3679 I \\
     -34.7319+26.0397 I & -39.4597+35.9027 I & 47.8881-14.0742 I & 56.949-22.7959 I \\
     59.5124+35.0029 I & 38.9296+65.799 I & -45.0976+9.69467 I & -48.8996-14.7589 I \\
     16.6759-64.4557 I & 12.8955-80.7994 I & 20.9437+39.2418 I & 30.1933+45.5707 I \
   \end{pmatrix} ^{\dagger} \\
    \begin{pmatrix}
     -16.0615-42.0538 I & 6.95806-49.3598 I & 0.380344+51.7601 I & -27.3546+50.3679 I \\
     -34.7319+26.0397 I & -39.4597+35.9027 I & 47.8881-14.0742 I & 56.949-22.7959 I \\
     59.5124+35.0029 I & 38.9296+65.799 I & -45.0976+9.69467 I & -48.8996-14.7589 I \\
     16.6759-64.4557 I & 12.8955-80.7994 I & 20.9437+39.2418 I & 30.1933+45.5707 I \
   \end{pmatrix} ) \\
   \; =  P_{un} ( \begin{pmatrix}
     -16.0615+42.0538 I & -34.7319-26.0397 I & 59.5124-35.0029 I & 16.6759+64.4557 I \\
     6.95806+49.3598 I & -39.4597-35.9027 I & 38.9296-65.799 I & 12.8955+80.7994 I \\
     0.380344-51.7601 I & 47.8881+14.0742 I & -45.0976-9.69467 I & 20.9437-39.2418 I \\
     -27.3546-50.3679 I & 56.949+22.7959 I & -48.8996+14.7589 I & 30.1933-45.5707 I \
   \end{pmatrix} \\
  \begin{pmatrix}
     -16.0615-42.0538 I & 6.95806-49.3598 I & 0.380344+51.7601 I & -27.3546+50.3679 I \\
     -34.7319+26.0397 I & -39.4597+35.9027 I & 47.8881-14.0742 I & 56.949-22.7959 I \\
     59.5124+35.0029 I & 38.9296+65.799 I & -45.0976+9.69467 I & -48.8996-14.7589 I \\
     16.6759-64.4557 I & 12.8955-80.7994 I & 20.9437+39.2418 I & 30.1933+45.5707 I \
   \end{pmatrix} ) \; = \\
P_{un} ( \begin{pmatrix}
     13110.4 & 14312.4+2902.95 I & -8737.18+2586.31 I & -10110.9+888.808 I \\
     14312.4-2902.95 I & 17870.7 & -8965.55+4758.04 I & -11909.6+3525.38 I \\
     -8737.18-2586.31 I & -8965.55-4758.04 I & 9276.95 & 10127.5+2054.13 I \\
     -10110.9-888.808 I & -11909.6-3525.38 I & 10127.5-2054.13 I & 12645.4 \\
   \end{pmatrix} ) \\
   = \; 26451.7 \; > \; 4
\end{multline*}
he bets on $ \sigma_{z} $.

 \item since:
\begin{equation*}
  \| a(3) \, - \, \sigma_{z} \| \; = \; \| \begin{pmatrix}
    8.80519-7.0523 I & -7.72367-6.40421 I \\
    -0.227234+7.87254 I & 7.36604+6.81784 I \
  \end{pmatrix} \| \; = \; 15.3175 \; > \; 10
\end{equation*}
he loses his fiche. Consequentially $ PAYOFF(2) \; = \; -1 $

\item since:
\begin{equation*}
  P_{un} ( \vec{a}(3)^{\dagger} \vec{a}(3) ) \; = \; 6.97591 \, 10^{6}
  \; > \; 8
\end{equation*}
he bets on $ \sigma_{z} $.

 \item since:
\begin{equation*}
  \| a(4) \, - \, \sigma_{z} \| \; = \; \| \begin{pmatrix}
    3.55982-1.58403 I & 2.19976-1.67009 I \\
    0.284886+2.77311 I & -7.06443-6.30601 I \
  \end{pmatrix} \| \; = \; 10.0665 \; > 10
\end{equation*}
he loses his fiche. Consequentially $ PAYOFF(3) \; = \; -2 $

\item since:
\begin{equation*}
  P_{un} ( \vec{a}(4)^{\dagger} \vec{a}(4) ) \; = \;  7.5079 \,
  10^{8} \; > \;  16
\end{equation*}
he bets on $ \sigma_{z} $.

\item since:
\begin{equation*}
   \| a(4) \, - \, \sigma_{z} \| \; = \; \| \begin{pmatrix}
     -8.49908+1.07129 I & -0.361299-7.07676 I \\
     9.60704+6.81686 I & -1.16288-3.10934 I \
   \end{pmatrix} \| \; = \; 14.1717 \; > \; 10
\end{equation*}
he loses his fiche. Consequentially $ PAYOFF(3) \; = \; -3 $
\end{itemize}

\bigskip

Exactly as it happened for the other kinds of Quantum Casinos,
the notion of a \textbf{third kind Quantum Casino} induces
naturally the notion of a \textbf{third kind collective}:

\begin{definition}
\end{definition}
THIRD KIND QUANTUM COLLECTIVES:

$ {\mathcal{Q}} {\mathcal{C}}ollectives^{3} \; \subset \;
\Sigma_{alg}^{\bigotimes \infty} $ induced by the
axiom\ref{ax:axiom of randomness} and the assumption that the
\textbf{third kind quantum admissible gambling strategies} are
nothing but the \textbf{quantum algorithms} on $
\Sigma_{alg}^{\bigotimes \infty}  $:
\begin{equation}
  {\mathcal{Q}}{{\mathcal{S}}trategies}_{admissible}
( {\mathcal{Q}} {\mathcal{C}}ollectives^{3} ) \; :=
{\mathcal{Q}}-{\mathcal{A}}lgorithms ( \Sigma_{alg}^{\bigotimes
\infty})
\end{equation}
\newpage
\section{The censorship of winning quantum gambling strategies}
Quantum Algoritmic Information Theory is a young field of
research in which there is not general agreement even on the
basic notion, i.e the correct way of defining quantum algorithmic
information, but a plethora of different attempts:
\begin{itemize}
  \item Karl Svozil's original creation of the research field \cite{Svozil-96}
  \item Yuri Manin's final remarks in his talk at the 1999's Bourbaki
  seminar \cite{Manin-99}
  \item Paul Vitanyi's definition \cite{Vitanyi-99}
  \item the definition by Andr\'{e} Berthiaume, Wim van Dam and
  Sophie Laplante \cite{Berthiaume-van-Dam-Laplante-00}
  \item our proposal \cite{Segre-00}
  \item the approach by Peter Gacs \cite{Gacs-00}
\end{itemize}

Whichever of these (or other new ones) attempts will appear to be
the right one, it will give rise to the construction of a whole
building at which last two floors there will be:
\begin{enumerate}
  \item the characterization of the notion of \textbf{quantum algorithmic
  randomness} as \textbf{quantum algorithmic incompressibility}
  \item the formulation and proof of \textbf{quantum-algorithmic-information undecidability
  theorems} analogous to Chaitin's Undecidability Theorems  \cite{Calude-96}
  poning constraints on the decidability of, respectively, quantum algorithmic
  information and the \textbf{quantum halting probability}
\footnote{May be they will finally result to be linked with the
quantum-logical violation of the \textit{Lindenbaum's property}
\cite{Dalla-Chiara-Giuntini-97}, \cite{Dalla-Chiara-Giuntini-01}?}
\cite{Svozil-96}
\end{enumerate}

Since effective-realizable measurements are particular
\textbf{quantum algorithms} the issue of characterizing  the right
notion of \textbf{quantum algorithmic randomness} is related with
the issue of the  \textbf{classical algorithmic randomness} of
quantum measuremnts' outcomes recentely analyzed By Ulvi
Yurtsever \cite{Yurtsever-98}.

Exactly as it happened for the classical notion of Martin L\"{o}f
Solovay Chaitin randomness, it is rather natural to think that
the right notion of \textbf{quantum algorithmic randomness} will
emerge as the more stable one, i.e. as that notion to which
completelly independent approaches belonging to completelly
different frameworks collapse to.

Clear, which ever it is, the right definition of  a random
quantum-sequence of qubits individuates the subset  $
RANDOM_{right}( \Sigma_{alg}^{\bigotimes \infty} ) $  of the
random sequences.

I think that the overwhelming majority of those  who has studied
Classical and Quantum Algorithmic Information Theory would bet on
the fact that, in the future,  someone  will show that $
P_{unbiased} ( RANDOM_{right}( \Sigma_{alg}^{\bigotimes \infty} )
) \; = \; 1 $.

Let us observe, by the way, that the theorem\ref{th:not existence
of Kolmogorov random sequences of cbits} doesn't generalize to
the quantum domain, since given a  \textbf{quantum probability
space} $QPA \; := \; ( \, A \, , \, \omega \, ) $ and introduced
the following straigthforward  noncommutative generalization of
the previously introduced classical notions:
\begin{definition}
\end{definition}
$ S \; \subset A $  IS A NULL SET OF QPS:
\begin{equation}
  E( a )  \; = \; 0 \; \; \forall a \in S
\end{equation}
\begin{definition}
\end{definition}
UNARY PREDICATES ON QPS :
\begin{equation}
  {\mathcal{P}} ( A ) \; \equiv \; \{ p( a ) \, : \, \text{ predicate about } a \in A  \}
\end{equation}
\begin{definition}
\end{definition}
TYPICAL PROPERTIES OF QPS:
\begin{equation}
 {\mathcal{P}}  (  QPS  )_{TYPICAL} \; \equiv \; \{ \, p ( a )  \in  {\mathcal{P}} ( A ) \, :
   \{ a \in A \, : \, p ( a ) \text{ doesn't hold } \}
   \; \text{is a null set} \}
\end{equation}
\begin{definition}
\end{definition}
SET OF THE  QUANTUM KOLMOGOROV  RANDOM ELEMENTS OF QPS:
\begin{multline}
  QUANTUM KOLMOGOROV-RANDOM ( QPS ) \; \equiv \\
   \{ \, a \, \in \, A \, : p ( a ) \; holds \; \; \forall p \in {\mathcal{P}}  (  QPS
  )_{TYPICAL} \; \}
\end{multline}
we have that the unbiased quantum probability of a single quantum
sequence is not necessary null, so that:
\begin{equation}
 \bar{a} \in \Sigma_{alg}^{\bigotimes \infty} \; \nRightarrow p_{\bar{a}} \notin  {\mathcal{P}}  (  UQS  )_{TYPICAL}
\end{equation}
where $  UQS \; := \; ( \, \Sigma_{alg}^{\bigotimes \infty} \, ,
\, P_{unbiased} \, ) $ is the \textbf{unbiased quantum probability
space of quantum sequences}, while $ p_{\bar{a}} $ is the
predicate $ p_{\bar{a}} ( \cdot ) \; := << \cdot \neq \bar{a}
>> $, implying  that the proof of the theorem\ref{th:not existence
of Kolmogorov random sequences of cbits} doesn't hold in the
quantum case.

Is $ RANDOM_{right}( \Sigma_{alg}^{\bigotimes \infty} ) \; = \;
QUANTUM - KOLMOGOROV-RANDOM ( UQS ) $?

It is highly probable that the answer to such a question is
negative, since the notion of quantum Kolmogorov randomness
doesn't seem to have the features of a notion candidated to be a
measure of (quantum) algorithmic incompressibility.

What we want to stress here is that, whichever $ RANDOM_{right}(
\Sigma_{alg}^{\bigotimes \infty} ) $ will be, its same meaning
requires that it satisfies the following constraint:
\begin{equation}
   {\mathcal{Q}} {\mathcal{C}}ollectives^{3} \; \subseteq \; RANDOM_{right}( \Sigma_{alg}^{\bigotimes \infty} )
\end{equation}

Can we give an assurance to a third kind Quantum Casinos' owner
that in the long run he doesn't risk anything?

The positive answer is stated by the following:
\begin{conjecture}
\end{conjecture}
LAW OF EXCLUDED QUANTUM GAMBLING STRATEGIES FOR THIRD KIND QUANTUM
CASINOS

For $ n \rightarrow \infty $ the set of  the \textit{lucky-winning
strategies} tends to the null set $ \forall \epsilon_{Casino} \in
{ \mathbb{R}}_{+} $.

\end{document}